\documentclass[12pt,preprint]{aastex}             
\slugcomment {\today} 
\usepackage{graphicx}
\usepackage{soul}

\shorttitle{}
\shortauthors{}
\begin{document}
\title{How to Identify and Separate Bright Galaxy Clusters from the Low-frequency Radio Sky?}
\author{Jingying Wang\altaffilmark{1}, Haiguang Xu\altaffilmark{1},
  Junhua Gu\altaffilmark{1},  Tao An\altaffilmark{2}, Haijuan Cui\altaffilmark{1},
 Jianxun Li\altaffilmark{3},  Zhongli Zhang\altaffilmark{4},
  Qian Zheng\altaffilmark{5},  Xiang-Ping  Wu\altaffilmark{5} }
 \altaffiltext{1}{Department of Physics, Shanghai
  Jiao Tong University, 800 Dongchuan Road, Minhang, Shanghai 200240,
  China; e-mail: hgxu@sjtu.edu.cn (Haiguang Xu), zishi@sjtu.edu.cn (Jingying Wang);}
\altaffiltext{2}{Shanghai Astronomical Observatory,
  Chinese Academy of Sciences, Shanghai 200030, China;}
 \altaffiltext{3}{School of Electronic Information and Electrical
  Engineering, Shanghai Jiao Tong University, 800 Dongchuan Road,
  Minhang, Shanghai 200240, China;} 
 \altaffiltext{4}{Max-Planck-Institute of Astrophysics,
  Karl-Schwarzschild-Str.1, Postfach 1317 85741 Garching, Munich,
  Germany;}  
  \altaffiltext{5}{National Astronomical Observatories, Chinese Academy
  of Sciences, Beijing 100012, China.}
 
\begin{abstract}
       
  In this work we simulate the $50-200$ MHz radio sky that is constrained in the field of view ($5^{\circ}$ radius) of the 21 Centimeter Array (21CMA), a low-frequency radio interferometric array constructed in the remote area of Xinjiang, China, by  carrying out Monte-Carlo simulations to model the strong contaminating  foreground of the redshifted cosmological reionization signals, including  emissions from  our Galaxy, galaxy clusters, and extragalactic discrete sources (i.e., star-forming galaxies, radio-quiet AGNs, and radio-loud AGNs). As an improvement of previous works, we consider in detail not only random variations of morphological and spectroscopic parameters within the ranges allowed by  multi-band observations, but also evolution of radio halos in galaxy clusters, assuming that relativistic electrons are re-accelerated  in the intra-cluster medium (ICM) in merger events and lose energy via both synchrotron emission and inverse Compton scattering with cosmic microwave background (CMB) photons.
By introducing a new approach designed on the basis of  independent component analysis (ICA)  and wavelet detection algorithm, we  prove that, with a cumulative observation of one month with the 21CMA array, about $80\%$ of galaxy clusters (37 out of 48 clusters  assuming a mean  magnetic field of $B=2~\mu$G in the ICM, or 15 out of 18 clusters assuming $B=0.2~\mu$G) with  central  brightness temperatures of  $> 10~{\rm K}$ at 65 MHz can be safely identified and separated from the overwhelmingly bright foreground.
By examining the  brightness temperature images and spectra extracted from these identified clusters, we find that the  morphological and spectroscopic distortions  are extremely small as compared to the input simulated clusters, and the reduced $\chi^2$ of brightness temperature profiles and spectra are controlled to be $\lesssim 0.5$ and $\lesssim 1.3$, respectively. These results robustly indicate that in the near future  a sample of dozens of bright galaxy clusters will be disentangled from the foreground in 21CMA observations, the study of which will greatly improve our knowledge about cluster merger rates, electron acceleration mechanisms in cluster radio halos, and magnetic field in the ICM.

\end{abstract}

\keywords{ early universe ---
 galaxies: clusters: general ---
radio continuum: general---
reionization---
 techniques: image processing
}

\section{INTRODUCTION}

Galaxy clusters are the largest virialized systems in the Universe.
Each of them contains about $10^{14}-10^{15}$ M$_{\odot}$ matter in the form of galaxies, hot gas  and dark matter, and all these components have been studied in depth in optical, X-ray, and high-frequency ($> 1$ GHz) radio bands (e.g., Sarazin \& Kempner 2000; Rosati et al. 2002; Popesso et al. 2004; Ferrari et al. 2008). However, the physical conditions of magnetic fields and high energy particles, which permeate the extremely hot intra-cluster medium (ICM) and have been playing important roles in the evolution of the ICM (e.g., Schekochihin et al. 2005; Lazarian 2006), are still poorly understood. In the study of  giant radio halos, which are always found  in  central regions of galaxy clusters and usually span on Mpc scales, these lacked knowledges  actually become crucial, because most of the emission of the radio halos can be ascribed to  synchrotron  radiation from relativistic electrons moving in the $\mu$G magnetic fields during or after a merger event (e.g., Ferrari et al. 2008; Brunetti et al. 2008). A variety of electron acceleration mechanisms and magnetic field distributions (e.g.,  En{\ss}lin \& Gopal-Krishna 2001; Brunetti et al. 2001, 2004; Cassano \& Brunetti 2005) have been proposed to construct dynamic models to describe  the  behavior of relativistic electrons, and these models are to be testified by the upcoming facilities working in the low-frequency ($< 1 ~{\rm GHz}$) radio band, such as the 21 Centimeter Array (21CMA;
Peterson et al. 2005), the Low Frequency Array (LOFAR; R\"{o}ttgering et
al. 2003), and the Murchison Wide-field Array (MWA; Morales \& Hewitt
2004). 

As of today, only $\lesssim 50$ detections of radio halos in the GHz radio band have been reported (e.g., Cassano et al. 2006; Giovannini et al. 2009; Brunetti et al. 2009). However, since the characteristic decay time of synchrotron emission becomes significantly longer when the frequency  goes lower (for example, the synchrotron decay time at 150 MHz is  approximately three times longer than that at 1.4 GHz), much more radio halos are expected  at frequencies of a few hundred MHz.  For example, Cassano et al. (2010)  estimated that more than 350 giant radio halos will be detected at 120 MHz by LOFAR in northern hemisphere and at Galactic latitudes $|b|\geq10^{\circ}$. And this expectation has been partially confirmed by recent Giant Metrewave Radio Telescope (GMRT) observations at 610 MHz (e.g., Venturi et al. 2007, 2008). Clearly, tight observational constrains on the formation and evolution of radio halos of galaxy clusters will be obtained in the next decade.

In most previous works  on  future low-frequency radio
 researches,   galaxy clusters  were only treated as  part of the useless
contaminating foreground  in front of the  redshifted 21 cm emission signals that come from
the cosmic reionization era, and  were simply filtered out, together with  the foreground emissions from our Galaxy and extragalactic discrete sources (e.g., Wang et al. 2006; Jeli\'{c} et al. 2008).   Little effort has been made to study the feasibility of separating the cluster component from the luminous foreground.   Meanwhile, we find that in the existing numerical works  models constructed for galaxy clusters are often oversimplified by either introducing
a universal spectral index for the radio flux density of  all
cluster halos ($\bar\alpha^{\rm CL}=1.2$; e.g., Gleser et al. 2008; Jeli\'{c} et al. 2008), or, sometimes, by
assuming a flat profile (e.g., Jeli\'{c} et al. 2008) for the spatial distributions of the radio surface
brightness. Similar oversimplifications  are often found in the simulations of the emissions from our
Galaxy and extragalactic discrete sources, too.

In order to construct the contaminating foreground  in front of the redshifted cosmic reionization signals  with reasonably higher spatial and spectral accuracies, by fully taking into account the random variations of model parameters in the ranges allowed by the observations, and  to investigate whether or not the cluster component (or, at least, brightest clusters) can
be successfully disentangled from the  bright foreground, we carry out
 Monte-Carlo simulations to model the Galactic,
cluster, discrete source, and the reionization  components, with which we create the $50-200$ MHz radio sky  with high fidelity as possible as we can. Then we
attempt to separate galaxy clusters and extract brightness temperature profiles and spectra  from the brightest clusters, whose central brightness temperature are brighter than $10~{\rm K}$ at $65~{\rm MHz}$, by introducing an approach designed on the basis of the independent component analysis (ICA; Comon 1994)
and wavelet detection algorithm (Shensa 1992).  All the  simulations are  performed for the 21CMA array that covers a  field of view (FOV) of about $5^{\circ}-10^{\circ}$ around the North Celestial Pole
(NCP), depending on the frequency.  Throughout the work we adopt $H_0$=71$h_{71}$ km s$^{-1}$
Mpc$^{-1}$, $\Omega_{\rm M}=0.27$, $\Omega_{\Lambda}=0.73$, and $\Omega_b=0.044$ (e.g., Spergel et al. 2003).

\section{SIMULATION OF THE LOW-FREQUENCY SKY}

In this section, we simulate the contributions that come from our Galaxy, galaxy
clusters, extragalactic discrete sources, cosmic reionization regions,
and antenna array, respectively.
 All the simulated results are plotted in a circular FOV with a $5^{\circ}$ radius around the NCP, as will be observed with the 21CMA array, although in our simulations we firstly create  corresponding images with a radius of $6^{\circ}$, so that we are able to correct the boundary effect. We project the sky field to a flat plane using the azimuthal equidistant projection, and the obtained images are overlaid with grids of $1024\times1024$ pixels, each pixel covering approximately a $0.6^{\prime} \times 0.6^{\prime}$ patch. 

\subsection{Our Galaxy}

In the low-frequency radio band the emission of our Galaxy is luminous
and  consists of synchrotron,
free--free, and dust components (e.g., Jones et al. 1998a; Platania et al. 2003; Gleser et al. 2008).  Of these, synchrotron emission is the dominating component,  and is responsible
for $\simeq 98.5\%$ of the total Galactic contamination in $100-200$
MHz. Free-free emission from the diffuse ionized hydrogen in the
interstellar medium (ISM)  is the next important contaminating source that contributes up to $\simeq 1.5\%$  of the total Galactic emission in $100-200$ MHz. Dust emission is mainly  restricted on the Galactic plane and is usually
negligible in $100-200$ MHz (Shaver et
al. 1999; Platania et al. 2003), thus we don't take it into account in this work.

{The diffuse Galactic synchrotron emission exhibits visible substructures in the total (Stokes I) intensity map, which can be attributed to the inhomogeneous spatial distributions of the magnetic field (e.g.,
Gaensler et al. 2001) and the ISM (e.g.,
Wieringa et al. 1993; Gray et al. 1999; Haverkorn et al. 2000). As of today, the smallest  substructures were resolved on the $\sim 1^{\prime}$ scales in the 1.4 GHz Southern Galactic Plane Survey (SGPS; Tucci et al.  2002), whereas  in the low-resolution Westerbork Telescope (WSRT) observations  $10^{\prime}$-scaled substructures at 150 MHz were reported as a common feature (Bernardi
et al. 2009).  
In our simulation, we construct the two-dimensional Galactic synchrotron map from observation results (Giardino et al. 2002) directly, rather than the modeling of three-dimensional Galactic structures  (e.g., Jeli\'{c} et al. 2008; Sun et al. 2008; Sun \& Reich 2009).
By adopting the 21CMA FOV corresponding region of the brightness temperature image obtained at 408 MHz ($T^{\rm Gsyn}_{408{\rm~ MHz}}({\bf r})$, where ${\bf r}$ is the two-dimensional position;  Giardino et al. 2002}), on which substructures with scales larger than $0.85^{\circ}$ are reproduced based on the 408 MHz all sky map (Haslam et al. 1982) and substructures  with smaller scales (but $\gtrsim 2^{\prime}$) are simulated using the extrapolation of the Galactic emission power-spectrum, we obtain the simulated Galactic synchrotron emission images $T^{\rm Gsyn}_{\nu}({\bf r})$ (where $\nu$ is the frequency;  see an example  $T^{\rm Gsyn}_{65~{\rm MHz}}({\bf r})$ in Fig.   1{\it a}) as
\begin{equation}
  T^{\rm Gsyn}_{\nu}({\bf r})=T^{\rm Gsyn}_{408~{\rm MHz}}({\bf r})(\frac {\nu }{408~{\rm MHz}})^{-\alpha_T^{\rm Gsyn}({\bf r})}, 
\end{equation}
where $\alpha_T^{\rm Gsyn}({\bf r}) \simeq 2.7-2.9$  is the corresponding spatially  varying synchrotron temperature spectral index\footnote[1]{According to Rayleigh--Jeans approximation, temperature spectral index $\alpha_T$ is related to the flux density spectral index $\alpha$ via $\alpha_T=\alpha+2$. } (Fig.   1{\it b}) calculated from the data of the 408 MHz and 1.4 GHz surveys (Giardino et al. 2002). Thus we obtain the two-dimensional
 Galactic synchrotron emission distribution $T^{\rm Gsyn}_{\nu}({\bf r})$,  which ranges from mean value of $5213$ K (root mean square, RMS hereafter, is $712$ K) at 50 MHz to $106$ K (RMS is $10$ K) at 200 MHz.

The Galactic free-free emission component cannot be directly
observed with today's technology, because the Galactic synchrotron component dominates the emission
at all frequencies lower than 10 GHz, meanwhile the dust thermal emission becomes overwhelming at higher frequencies. However, hydrogen recombination
lines ($H\alpha$ line emission in particular) provide a novel way to trace the free-free
emission, since they originate in nearly the same region (Smoot 1998; Reynolds \& Haffner 2000). In fact, the
brightness temperature distribution of the Galactic free-free emission component at 30 GHz $T^{\rm Gff}_{30{\rm ~GHz}}({\bf r})$
 has been related to the $H\alpha$ intensity distribution $I_{H\alpha}({\bf r})$ (Finkbeiner 2003)  in the form of 
\begin{equation}
T^{\rm Gff}_{30{\rm ~GHz}}({\bf r})=7.4\frac{I_{H\alpha}({\bf r})}{\rm Rayliegh}~~{\rm \mu K},
\end{equation}
where ${\rm 1~Rayliegh\equiv 10^6/4\pi~photons~ s^{-1} ~cm^{-2}~
  sr^{-1}}$ (Reynolds \& Haffner 2000).  Using this relation, we derive the $T^{\rm Gff}_{30{\rm ~GHz}}({\bf r})$ map of the 21CMA FOV from the corresponding region of $I_{H\alpha}({\bf r})$ survey image (Finkbeiner 2003). By assuming that the brightness temperature spectrum of the Galactic free-free emission follows a broken power-law shape,  i.e., $T^{\rm Gff}_{\nu}
\propto \nu^{-\alpha_{T}^{\rm Gff}}$, where $\alpha_T^{\rm Gff}=2.10$ at $\nu \leq 10~{\rm GHz}$ (Shaver 1999) or $\alpha_T^{\rm   Gff}=2.15$ at $\nu >{\rm 10~GHz}$ (Bennett et al. 2003) is the free-free temperature spectral index, we theoretically obtain the  two-dimensional Galactic free-free emission distribution $T^{\rm Gff}_{\nu}({\bf r})$ (see an example  $T^{\rm Gff}_{65~{\rm MHz}}({\bf r})$ in Fig.   1{\it c}),  which ranges from mean value of $9$ K (RMS is $2$ K) at 50 MHz to $0.5$ K (RMS is $0.1$ K) at 200 MHz.

\subsection{Galaxy Clusters}
\subsubsection{Mass Function and Three-dimensional Spatial Distributions}
The mass distribution of dark matter halos at a given redshift $z$ can
be derived in the frame of the Press-Schechter  model (P-S model; Press \&
Schechter 1974), which has been extensively  used in a variety of works (e.g., Zhang \& Wu 2003; Zentner et al. 2005; Fedeli et al. 2006) since it was proposed.
According to the P-S model, the initial perturbation of dark matter density at a given position ${\bf R}$, i.e., $\rho({\bf R})$, is characterized by the density contrast $
\delta({\bf R})=\frac{\rho({\bf R})-\left<\rho\right>}{\left<\rho\right>} $ with its probability  $p(\delta)$ following
the Gaussian distribution  $ p(\delta)=\frac
{1}{\sqrt{2\pi}\sigma(M)}~e^{-\delta^2/2\sigma^2(M)}, $ where $\sigma(M)$ is the variance in the density field and $M$ is the mass of dark halo.  For a smoothed
density field with a constant  smoothing scale of $R_{sm}$, inside which the total mass is $M$, the variance in the density field is $
\sigma(M)=\left<\delta^2_{R_{sm}}({\bf R})\right>=\frac
{1}{2\pi^2}~\int{W^2_{R_{sm}}(k)~P(k)~k^2~dk}, $ where  $k$ is the wave number,
$W_{R_{sm}}(k)=e^{-k^2 R_{sm}^2/2}$ is the filter function for Gaussian
distribution, and $P(k)=|\delta_k|^2$ is the primordial power-spectrum. Based on
these the dark halo mass function can be expressed as
\begin{equation}
dn(M)=-\sqrt\frac{2}{\pi}~\frac{\left< \rho \right>}{M}~\frac{\delta_{c}}{\sigma^2(M)}~\frac{d\sigma(M)}{dM}~\exp\left(-\frac{\delta^2_c}{2\sigma^2(M)}\right)~dM,
\end{equation}
where $\delta_c(z)=\frac {\delta_c(0)}{D(z)}$ is the halo collapse
time, $D(z)$ is the normalized growth factor, and $\delta_{c}(0)=
0.15~(12\pi)^{2/3}~\Omega_{\rm M}^{0.0055}$, in which $\Omega_{\rm M}$ is
the total matter density. 

 Assuming a mass fraction of 83\% for dark
matter (e.g., Drexler 2007) and a  lower mass
limit of $2\times 10^{14}~ {\rm M_{\odot}}$ (e.g., Holder et al. 2001;
Voit 2005) for galaxy clusters, we calculate the number density distribution of galaxy clusters as a function of mass and redshift as  $n^{\rm CL}=n(M^{\rm CL}, z)$ ($M^{\rm CL}$ is the cluster mass)  according to the P-S model, and the redshift distribution of galaxy clusters as $n^{\rm CL}=n^{\rm CL}(z)$  (Fig. 2).  By examining the
derived $n^{\rm CL}(z)$, we find that galaxy clusters  mainly
distribute in the redshift range of $0<z<3$ (e.g., Holder et
al. 2001), which agrees with the observations, and 1084 cluster-level halos are expected in the 21CMA FOV.  In the work that follows we assign random $M^{\rm CL}$ and $z$ to these 1084 clusters according to the  
$n(M^{\rm CL}, z)$ distribution derived above, and determine the coordinates of the clusters  on the sky plane (i.e., R.A. and
Dec)   according to the predictions of the theoretical cluster power spectrum in redshift range of  $0<z<3$ (Majumdar \& Mohr 2004).

\subsubsection{Spectral Model}

Galaxy clusters that contain giant, diffuse radio halos are observed to exhibit a strong
correlation between their X-ray and radio powers (e.g., En{\ss}lin \&  R\"{o}ttgering 2002), whereas
those contain no or weak radio halos are found to disobey this
relation, showing a significantly lower $L_{\rm radio}/L_{\rm X}$ ratio (e.g., Cassano 2009), where  $L_{\rm radio}$ and $L_{\rm X}$ are the radio and X-ray luminosities for the clusters, respectively. According to the standard turbulent re-acceleration model, the primordial relativistic electron population accelerated in a merger event follows a power-law energy
distribution $N(\gamma,t)|_{t=0} \propto \gamma_0^{-\Gamma}$, where the Lorentz factor is $\gamma \gg 1000$, $\gamma_0=\gamma|_{t=0}$, and  the electron energy spectral index  is $\Gamma \gtrsim
3$ (e.g., Ferrari et al. 2008). These electrons lose energy not only via synchrotron emission when moving in
the  magnetic field ($B$), but also via inverse Compton
scattering through interacting with the cosmic microwave background (CMB) photons. The latter process can
be described in a similar way to the synchrotron emission if an equivalent magnetic field of $B_{\rm CMB}=3.2 (1+z)^2 ~{\mu}$G is introduced (e.g., Parma
et al. 1999), thus  the electrons lose their
energy on a time scale of
\begin{equation}
  \tau \approx 2\times10^{3}~\gamma^{-1}~\left[(1+z)^4+(\frac {B} {3.2~\mu{\rm G}})^2\right]^{-1}~~\rm{Gyr}
\end{equation}
(Ferrari et al. 2008).
 Taking $z=0.2$ (around which most of the giant radio halos are found) and the typical magnetic field strength  $B= 2~\mu$G (e.g., Kim et al. 1990) or $0.2~\mu$G (e.g., Ferrari et al. 2008; Rephaeli \& Gruber 2003; Rephaeli et al. 2006) for example, we find that the emission of the radio halo at 1.4 GHz will fade away after  $\tau \simeq 0.06~{\rm Gyr}$ or $\simeq 0.02$ Gyr for $B=2~\mu$G  or $0.2~\mu$G, respectively. At 100 MHz, the corresponding decay time will increase significantly to $\tau \simeq 0.2 ~{\rm Gyr}$ or  $\simeq 0.1$ Gyr for $B=2~\mu$G  or $0.2~\mu$G, respectively.
This means that the spectrum of  a radio halo will evolve into a soft state quickly, and finally fade away in the interval between two merger events (typically 2.7 Gyr; Mathiesen
\& Evrard 2001).  As a result, typical radio temperature spectral indices of galaxy clusters are $\alpha_{{T},~\nu}^{\rm CL}>3$ (Fig.   3{\it a}), which are significantly larger than those of other foreground components, including our Galaxy (\S 2.1) and extragalactic discrete sources (\S 2.3). In this work, we assign a random evolving time $t_E$ between 0 and 2.7 Gyr for each cluster, and calculate the low-frequency flux density $F_{\nu}^{\rm CL} (t_E)$ from its power density $P^{\rm CL}_{\nu}(t_E)$ (see details in Appendix) by assuming that  each cluster owns a radio halo produced in the last merger and the initial 1.4 GHz power density $P^{\rm CL}_{\rm 1.4~GHz}(t=0)$ obey the observed correlation between the 1.4 GHz power density and X-ray temperature $T_{\rm X}$.  In order to derive the approximation of $P^{\rm CL}_{1.4~ {\rm GHz}}-T_{\rm X}$ relation, using the data from Cassano et al. (2006), we repeatedly fit the observed data with maximum likelihood method  and exclude the clusters fall out of the $3\sigma$ lower limit, as is shown in Figure 3{\it b}. Then we have
\begin{equation}
P^{\rm CL}_{1.4~ {\rm GHz}}(t=0)=1.04\times10^{29} \left(\frac {T_{\rm X}} { \rm keV}\right)^{2.88} ~{\rm ergs~s^{-1}~Hz^{-1}}.
\end{equation}
 As a comparison,  on the figure we also superpose the data of three new cluster samples studied
by Giovannini et al. (2009), and find that all of them locate below the line described by equation (5).  The X-ray temperatures of the clusters are 
deduced from the mass-temperature relation provided by Arnaud
et al. (2005), i.e.,
\begin{equation}
T_{\rm x} =5~\left[ \frac {h(z)M^{\rm CL}} {5.34 \times 10^{14} M_{\odot}}\right]^{1/1.72} ~{\rm keV},
\end{equation}
where $h(z)=\sqrt{\Omega_{\rm M}(1+z)^3+1-\Omega_{\rm M}}$ is the
Hubble constant normalized to its local value.

\subsubsection{Morphologies}

Since for galaxy clusters there exists a linear correlation between the radio and X-ray
surface brightness  distributions (e.g., Govoni et
al. 2001), we simulate the radio surface brightness density distribution by applying the $\beta$ model
(Cavaliere \& Fusco-Femiano 1976)
\begin{equation}  
 S^{\rm CL}_{\nu}(r)= S_{0,~\nu}^{\rm CL}\left[1+\left(\frac r{r_c}\right)^2\right]^{-3\beta+0.5},
\end{equation}
where $r_c$ is the core radius, $\beta$ is the index, and
$S_{0}^{\rm CL}(\nu)$ is the central surface brightness density.  The
core radius $r_c$ is derived from the cluster's X-ray luminosity in
$0.5-2.0$ keV ($L_{\rm X,0.5-2.0keV}$) by 
\begin{equation}
r_c=\frac {176}{h_{71}}\left(\frac { L_{\rm X,0.5-2.0keV}}{5\times 10^{44} ~\rm ergs~s^{-1}}\right)^{0.2}~{\rm kpc}
\end{equation}
(Jones et al. 1998b), and $ L_{\rm X,0.5-2.0keV}$ is calculated from the bolometric
($0-20$ keV) luminosity using
\begin{equation}  
{ L_{\rm X, bolo}}=12.44\pm1.08 \times10^{44}~(\frac { T_{\rm X}}{6\mbox{ keV}})^{2.64\pm0.27}~ (1+z)^{1.52^{+0.26}_{-0.27} } ~~~{\rm ergs~ s}^{-1}
\end{equation}
(Lumb et al. 2004),  assuming the metal abundance as a function of redshift as given in Ettori (2005). On the other hand, we calculate the parameter $\beta$ by applying equation (2) of Evrard et al. (1996)
\begin{equation}
\beta=8.85\times10^{-15}~ {\frac {1+(r_{200}/r_c)^2} {(r_{200}/r_c)^2}}~\left({\frac {T_{\rm X}} {\rm keV}}\right)^{-1}~\left({\frac {r_{200}} {\rm Mpc}}\right)^{-1}~{\frac {M^{\rm CL}} {M_{\odot}}}, 
\end{equation}
where $r_{200}$ is the virial radius within which the mean interior over-density is 200
times of the cosmic critical density at the redshift considered. 
When $r_c$ and $\beta$ are determined, $S_{0,~\nu}^{\rm CL}$ can be constrained from cluster's flux density $F^{\rm CL}_{\nu}(t_E)$, which is given in $\S{\it 2.2.2}$, by
 \begin{equation}
{ F^{\rm CL}_{\nu}(t_E)}=\int^{r_{200}}_0 2\pi r { S_{0,~\nu}^{\rm CL}}\left[1+\left(\frac r {r_c}\right)^2\right]^{-3\beta+0.5}dr.
\end{equation}

Thus, having derived the parameters $r_{200}$, coordinates, $r_c$, $\beta$, and $S_{0,~\nu}^{\rm CL}$ of each cluster,  we generate the two-dimensional surface brightness density map of each galaxy cluster at $\nu$ using equation (7), overlay all the maps together to construct the simulated map of cluster component in the 21CMA FOV, and convert the map into simulated cluster brightness temperature image $T_{{\rm B},\nu}^{\rm CL}({\bf r})$ (see examples at 65 MHz  in Fig.   4) with the Rayleigh--Jeans approximation
\begin{equation}
T_{{\rm B},\nu}^{\rm CL}({\bf r})=\left( {\frac {2\nu^2} {c^2} } k_{\rm B} \right)^{-1} S^{\rm CL}_{\nu}({\bf r}).
\end{equation}

\subsection{Extragalactic Discrete Sources}

In most simulations on the foreground of the redshifted 21 cm reionization signals (e.g., Wang et
al. 2006;  Santos et al. 2005; Jeli\'c et
al. 2008), it is assumed that all the extragalactic discrete sources are
point-like and possess the same radio spectrum in the form of $F_{\nu}\propto
\nu^{-0.7}$, whereas the observations indicate a more complicated situation. Following the works of Wilman et
al. (2008; W08 hereafter) and Snellen et al. (2000; S00 hereafter), we classify the extragalactic sources into four
types, i.e., 1) star-forming galaxies, including normal star-forming galaxies and
starburst galaxies; 2) radio-quiet AGNs; 3) Fanaroff-Riley class I (FRI) and class II (FRII)
AGNs; and 4) GHz peaked spectrum (GPS) and compact
steep spectrum (CSS) AGNs. We simulate the flux densities, spatial structures, spectra, and angular
clusterings of types (1)-(3) sources following Wilman et al. (2008) and references therein, and those of type (4) sources  following Snellen et al. (2000) and reference therein. 
To be specific, we start with the flux density functions that were obtained for types (1)-(3) sources at a fiducial frequency $\nu_f=1.4$ GHz (Fig. 4 of W08), and the evolving luminosity function for type (4) sources at $\nu_f=5$ GHz (Eq. (21) of S00), respectively, to derive the flux densities of the sources that should appear in the 21CMA FOV, and simulate their spatial structures to create the brightness temperature images at $\nu_f$. Then with the information of their spectral properties, we simulate the brightness temperature images in $50-200$ MHz band for each type of extragalactic discrete sources (see  examples at 65 MHz in Fig. 5), and obtain the brightness temperature images  of the extragalactic discrete component  $T_{\rm B,~\nu}^{\rm  Disc}({\bf r})$. The simulation details in our process are as follows.

\noindent {\bf Star-forming galaxies}\\
\indent (i) Each source is assumed to be disk-like and located on the sky plane with a random orientation. For a normal star-forming galaxy, the linear size $D$ is related to the 1.4 GHz luminosity by $\log (\frac {D}{\rm kpc})=0.22\log (\frac {L_{1.4 {\rm ~GHz}}} {\rm W ~Hz^{-1}~sr^{-1}})-\log(1+z)-3.32(\pm0.18)$ (Eqs. (7)-(9) of W08). For a starburst galaxy, we assume that $D=(1+z)^{2.5}~{\rm kpc}$ (Eq. (10) of W08) out to $z=1.5$ and $D \equiv 10~{\rm kpc}$ for $z>1.5$.\\
\indent (ii) The emission of each star-forming galaxy consists of one thermal free-free and one non-thermal synchrotron component, which are constrained by the ratio of the total luminosity to the luminosity of the thermal component of $1+10\left({\frac {\nu} {\rm GHz}}\right)^{-0.65\pm0.10}$ (W08). \\
\indent (iii) The spectrum of each source is modeled with two power-laws as
$L_{\nu}\propto \left({\frac {\nu} {\rm GHz}}\right)^{-0.10}+10 \left({\frac {\nu} {\rm GHz}}\right)^{-0.75\pm0.10}$ (Eq. (6) of W08),
where the two terms represent the thermal free-free and
non-thermal synchrotron components, respectively.

\noindent {\bf Radio-quiet AGNs}\\
\indent (i) Sources are all observationally point-like.\\
\indent (ii) The spectrum of each source can be  modeled  with a power-law model as $F_{\nu}\propto \nu^{-0.7}$ (W08) that is normalized by using the 1.4 GHz luminosity.

\noindent{\bf FRI \& FRII AGNs}\\
\indent (i) Each source is assigned with a linear size that follows a random, uniform distribution $\left[0, D_0(1 + z)^{-1.4}\right]$ at a given redshift, where $D_0 = 1$ Mpc. Each source possesses a point-like compact core, and two extended lobes that can be modeled as ellipses with the axial ratio  drawn from a uniform distribution $\left[0.2,1\right]$ (W08).\\
\indent (ii) According to the orientation-based unification model, the ratio of the 1.4 GHz flux density of the core to that of the lobes is $F^{\rm core}_{\rm 1.4~GHz}/F^{\rm lobe}_{\rm
  1.4~GHz}=R_{\rm rf} B(\theta)$, where $R_{\rm rf}=10^{x}$ is the rest-frame
ratio with $x$ following a Gaussian distribution $N(\bar{x},0.5)$, and $B(\theta)=1/2[(1-\sqrt{(\gamma
  ^{2}-1)/\gamma}\cos\theta)^{-2}+(1+ \sqrt{(\gamma
  ^{2}-1)/\gamma}\cos\theta)^{-2}]$ with $\gamma$ being the Lorentz
factor of the jet and $\theta$ being the angle between jet axis and
 line-of-sight (i.e., inclination angle) that distributes uniformly between $0$
and $\pi$. We set $(\bar{x},\gamma)=(-2.6,6)$ and $(-2.8,8)$ for FRI and FRII sources, respectively (W08).\\
\indent (iii) The spectrum of the core emission is estimated as
${\rm \log}F^{\rm core}_{\nu}=a_{0}+a_{1}{\rm \log}(\frac {\nu}{\rm GHz})+a_{2}{\rm
  \log}^{2}(\frac {\nu}{\rm GHz})$, with $a_{1}=0.07\pm0.08$, $a_{2}=-0.29\pm0.06$ and $a_0$ being calibrated by applying $F^{\rm core}_{\rm 1.4~GHz}$.
The spectrum of the extended lobe emission is parameterized as $F^{\rm lobe}_{\nu}\propto
\nu^{-0.75}$ (W08).

\noindent {\bf GPS \& CSS AGNs}\\
\indent (i) Sources are all observationally point-like.\\
\indent (ii) The spectral shape is found to have a turnover below several GHz, which can be described as
\begin{equation}
F^{\rm GPS,CSS}_{\nu}=\frac{F^{\rm GPS,CSS}_{\nu_{p}}}{1-e^{-1}}(\frac{\nu}{\nu_{p}})^{t_k}\left[1-e^{-(\frac{\nu}{\nu_{p}})^{t_n-t_k}}\right]
\end{equation}
(Eq. (1) of Snellen et al. 1998), where $t_k=0.51\pm0.03$ and $t_n=-0.73\pm0.06$ are the optically thick and
thin spectrum indices (de Vries et al. 1997), respectively, and $\nu_{p}$ is the observed
turnover frequency. The turnover frequency varies with 
$\log\left[\frac {\nu_{p}~(1+z)} {\rm GHz
  }\right]=-0.21(\pm0.05)-0.65(\pm0.05)\log(\frac {D} {\rm kpc})$ (Eq. (4) of W08), in
which the projected linear size of the source $D$ is independent from redshift but
follows the scale function between $D$ and the source number $N$,
\begin{equation}
dN/dD\propto\left\{
\begin{array}{ll}
 D^{-1} ~~~~~~~~~~~~ 0.01~{\rm kpc}\le D\le 0.355h_{71}^{-1}~{\rm kpc}\\
 D^{-0.6} ~~~~~~~~~~ 0.355h_{71}^{-1}~{\rm kpc}<D \le 20~~{\rm kpc}\\
\end{array}
\right.
\end{equation}
(Fanti et al. 2001; O'Dea 1998; Stawarz et al. 2008).

The total flux density functions of our simulated extragalactic discrete sources are shown in Figure 6.

\subsection{ Effects of Instrument}
The 21CMA array is a low-frequency radio interferometer  array constructed in a remote area of Xinjiang, China. The array consists of 81 antenna pods that are distributed in a ``T'' shape with the north-south and east-west baselines are 6 km and 4 km, respectively, and each pod is an assembly of 127 logarithmic periodic antennas. The array is operated in 50-200 MHz with the minimum bandwidth of 49 kHz. 

In radio interferometric observations, the systemic uncertainty for
one image pixel can be expressed in terms of brightness temperature as
\begin{equation}
\sigma_{\rm pix}=T_{\rm sys} \frac {\lambda^2} {A_{\rm eff} \Omega_{\rm beam} \sqrt{\Delta \nu \tau n (n-1)}}
\end{equation}
(Thompson et al. 2001), where $T_{\rm sys}$ is the system temperature,  $\lambda$
is the wave length, $A_{\rm eff}$ is the 
effective area of one antenna pod, $\Omega_{\rm beam}$ is the beam solid angle of the interferometric array, $\Delta\nu$ is the
bandwidth of one frequency channel, $\tau$ is the integral time,  and $n$ is the number of antenna pods. The 21CMA beam varies with the wavelength $\lambda$, e.g., the full width at half maximum (FWHM) of the beam is about $3.2^{\prime}$ (corresponding to $\Omega_{\rm beam}=6.8\times 10^{-7} ~{\rm
  sr}$) at 65 MHz, and  about $1.6^{\prime}$ ($\Omega_{\rm beam}=1.7\times 10^{-7} ~{\rm
  sr}$) at 150 MHz. For 21CMA, we have $T_{\rm sys}=300~ {\rm K}$,
$A_{\rm eff}=216 ~{\rm m^2}$,  $n=81$, thus, 
\begin{equation}
\sigma_{\rm pix}=252 {\rm~ mK}\left(\frac{\lambda}{2{~\rm m}}\right)^2\left(\frac{1 \rm{~ MHz}}{\Delta \nu}\right)^{1/2}\left(\frac{30 \rm{~days}}{\tau}\right)^{1/2}\left(\frac{1.7 \times 10^{-7}\rm{~sr}}{\Omega_{\rm beam}}\right).
\end{equation}
 Supposing a typical survey of $\tau= 30$ days and $\Delta \nu=1 ~{\rm MHz}$, for each frequency channel we generate the instrument noise as white noise and add it into the simulated 21CMA map.

By mixing the simulated brightness temperature images of all celestial and noise components, we create the final sky maps constrained in the 21CMA FOV (${T_{\rm B,~\nu}^{\rm Total}}({\bf r})$, where $50$ MHz $\leq\nu\leq200$ MHz, $\Delta\nu=1$ MHz), assuming that in the ICM of galaxy clusters there exists a uniform magnetic field $B=2 ~\mu$G and $0.2~\mu$G, respectively.
 We apply a  universal Gaussian kernel that is required by following component separation approach (see details in \S3), whose FWHM$=3.2^{\prime}$ corresponds to the beam solid angle $\Omega_{\rm beam}=6.8\times 10^{-7} ~{\rm
  sr}$ at $65$ MHz, to approximate the point spread function (PSF) of the observation, and convolve the simulated sky maps that have been cumulated within selected frequency channels  (central frequency $\nu=61,~62,~ 63,~...,~ 189$ MHz, respectively) to have a channel bandwidth of $\Delta\nu=1$ MHz for each. These fake 21CMA survey maps are marked as ${\bar{T}_{\rm B,~\nu}^{\rm Total}}({\bf r})$ hereafter (see  examples with the central frequency of $\nu=65$ MHz in Fig. 7$a$ \& 7$b$).

For the convolved sky maps, we calculate the confusion level with three different methods, and have crosschecked the results to confirm if they are consistent with each other (Table 2). In the first method, the confusion level is defined in a traditional way in terms of source flux density, over which there is 0.1 source per synthesized beam on average (e.g., R\"{o}ttgering et al. 2006). We adopt the source flux density function ($N(>F_{\nu})$) that derived from our simulation (\S2.3 and Fig. 6) and the beam  (FWHM=$3.2^{\prime}$) of our simulated survey maps to calculate the confusion level at 65 MHz and 150 MHz. In an alternative way, we also replace our source flux density function with the literature source flux density function that has been used to evaluate the confusion limit of LOFAR (Appendix B and Figure 4 of R\"{o}ttgering et al. 2008). In the second method, we employ the photometric criterion for confusion level as mentioned in Dole et al. (2003). From their equations (5)-(6) and related description, we deduce the confusion level to be $[(3+a)/(\Omega_{\rm beam}F_0q^2)]^{1/(a+1)}$, where $a$ and $F_0$ are the index and normalization of the differential flux density function ($-dN / dF_{\nu}$), respectively, and $q=3$ is the minimum signal to noise ratio for source detection.  In the third method,  we directly detect the sources on our simulated survey maps with the tool SExtractor (Bertin \& Arnouts 1996) at a $3\sigma$ confidence level. We derive the confusion level by counting the source number as a function of flux density, which is the approach employed by V{\"a}is{\"a}nen et al. (2001).  Meanwhile, we also calculate the confusion level with the instrument beam size of the 21CMA array at 65 MHz (FWHM=$3.2^{\prime}$ also) and 150 MHz (FWHM=$1.6^{\prime}$), respectively, by repeating the steps above. Since all the calculations above are at a $3\sigma$ confidence level, the value of confusion noise should be 1/4 of the confusion level.

\section {IDENTIFICATION OF BRIGHT GALAXY CLUSTERS}

In this section, after extremely bright  discrete sources are masked, we apply the ICA separation and wavelet detection algorithms to these maps to identify and study bright galaxy clusters, as described below.

\subsection{Step 1: Masking Extremely Bright Discrete Sources}
According to our simulations, except for our Galaxy, extragalactic discrete sources contribute most of the 
foreground in $50-200$ MHz in the 21CMA FOV. Of these, about $60$ sources are
extremely bright, and should be discarded before identifying and 
 analyzing weaker components. Here we introduce a masking method, which has been widely used in the CMB separation works (e.g., Bennett et
al. 2003; Leach et al. 2008), to exclude  map pixels that contain
extremely bright discrete sources. To be specific, we adopt a brightness temperature criterion of $\bar{T}_{\rm B,~\nu}^{\rm Total}({\bf r})>1.5 \times 10^4~{\rm  K}$ in all selected frequency channels to exclude circular regions with a radius of one FWHM ($3.2^{\prime}$) centered at all  extremely bright sources. In the analysis that follows, if
a pixel is masked in any one frequency channel, it is also masked in all other
channels. As a result, about $0.59\%$ of the 21CMA FOV is masked.

\subsection{Step 2: ICA Separation and Wavelet Detection}

\subsubsection{Raw Separation with FastICA}

As a blind signal separation method, ICA only assumes that the maps of original celestial and noise  components are independent from each other in frequency space, and  no more than one of them is Gaussian. These assumptions are applicable to our case. 
Thus, in terms of the ICA algorithm, the relation between the mixed maps obtained in different frequency channels, all of which have been whitened (Hyv\"{a}rinen 1999) as a pre-procession, and original component maps can be expressed as
\begin{equation}
{\bf x}={\bf As},
\end{equation}
where ${\bf x}=\{x_1,x_2,...,x_m\}^T$ is the vector constructed by  the mixed maps at $m$ layers of frequency channels ($i=1,~,2~,...,~m$), ${\bf s}=\{s_1,s_2,...,s_n\}^T$ is the vector constructed by $n$ ($n$ is the number of celestial components, $n \leq m$) layers of original celestial maps, and ${\bf A}$ is an $m \times n$ dimensional transition  matrix. Since from equation (17) we have  
\begin{equation}
{\bf s}={\bf A}^{-1}{\bf x},
\end{equation}
where ${\bf A}^{-1}$ is the inverse of ${\bf A}$, the original celestial maps can be derived when  ${\bf A}^{-1}$ is available. In order to determine  ${\bf A}^{-1}$, we introduce a random Gaussian field $\mu$,  with zero mean and unit variance, which has the same image size as the fake 21CMA maps, and  maximize the difference between two fields $\mu$ and ${\bf w}_i^T{\bf x}$,   where  ${\bf w}_i$ is  the $i$th ($i=1,~,2~,...,~n$) row of the test matrix ${\bf W}$  with unit norm,   by maximizing the approximation of nongaussianity,
\begin{equation}
J_G({\bf W})=\sum_{i=1}^n\left\{E\left[G({\bf w}_i^T{\bf x})\right]-E\left[G(\mu)\right]\right\}^2
\end{equation}
(Hyv\"{a}rinen 1999),  where 
$G$ is the nonquadratic operator for $\log\cosh$, and the expectation $E$ are calculated over the whole image. When the maximum is found, we have ${\bf W}={\bf A}^{-1}$.

As shown in equation (17), the ICA algorithm reads in and treats the input mixed maps as vectors, the elements of which are related to the pixels of the corresponding maps. The way to establish the mapping between a map and its corresponding vector (e.g., row-by-row, column-by-column, or even randomly) is arbitrary, but once determined for one map, it should keep the same for all the maps. Thus the spatial relations among pixels (i.e., image domain features) are broken and are not considered by the ICA algorithm. This is different from the traditional way to detect discrete sources on a map, for which the confusion noise is essentially important.

ICA has been proved to be an
effective separation technique in astronomical imaging analysis works, to reconstruct the sources confused by other components or even sunk into the confusion noise. In De Bernardis et al. (2009; the B-Pol proposal), the authors proposed to adopt ICA technique to extract dim point sources, which cannot be directly detected from the residual confusion noise but would dominate over the CMB B-mode signal.  
Baccigalupi et al. (2000) applied the ICA technique to a
$15^{\circ}\times15^{\circ}$ sky patch at high Galactic latitude ($l\geq82.5^{\circ}$), and successfully reconstructed both the CMB signals with an RMS error of about 1\%, and the
weaker (about two orders of magnitude fainter) Galactic components with an accuracy of $\sim 10\%$.
 Maino et al. (2006) and Bottino et al. (2008) applied an optimized ICA (FastICA; Hyv\"{a}rinen 1999)  technique, which adopts fixed-point algorithm in ICA to maximizing $J_G$ more efficiently, on the whole sky Wilkinson Microwave Anisotropy Probe (WMAP) map to extract the CMB signals and separate the multiple foreground components. 

In this work we attempt to use the FastICA package embedded in the Modular Toolkit for Data
Processing (MDP; Zito et al. 2009) to disentangle the cluster component from the fake
 21CMA survey maps. 
To reduce the complexity of the calculation, we divide the 21CMA FOV into
one circular and 24 pie patches (Fig. 7$c$ \& 7$d$) that have the same areas.  In order to correct the boundary effect, we enlarge each
patch by at least $10^{\prime}$ so that it slightly overlaps all the adjacent patches.   We carry out separation calculation in each patch by feeding the fake 21CMA  maps ${\bar{T}_{\rm B,~\nu}^{\rm Total}}({\bf r})$  obtained in the $\nu=$ 65, 80, 170, and 185 MHz channels into the FastICA program.  In the calculation, we assume that each celestial source bears similar  morphologies  in different frequency channels, due to the source correlation coefficient between any two channels is $> 0.9$, which is required by the algorithm. We start the FastICA calculation in Patch I with an  arbitrarily chosen test matrix ${\bf W}_{\rm I}^0$. When the maximum $J_G$ is found as ${\bf W}_{\rm I}^{\rm max}(={\bf A}_{\rm I}^{-1})$, we feed  ${\bf W}_{\rm I}^{\rm max}$ into the FastICA calculation of Patch II as the original test matrix ${\bf W}_{\rm II}^{0}$, because this will greatly reduce the iteration steps to find ${\bf W}_{\rm II}^{\rm max}(={\bf A}_{\rm II}^{-1})$. After repeating this progress for all patches to find all ${\bf W}_i^{\rm max}(={\bf A}_{\rm i}^{-1}; i={\rm I,~ II,~ III,~ ...,~ XXV})$, we obtain raw separated images $\bar{T}{_{\rm B,~\nu}^{\rm Src}}^{ \prime}({\bf r})$  (Src=Gsyn, Gff, CL, and Disc for Galactic synchrotron, Galactic free-free, galaxy cluster, and discrete source components, respectively) using equation (18). 

In each frequency channel the output raw separated images for different components are rescaled to guarantee that the cumulative brightness temperature at each image pixel equals to the corresponding input value.
 To deal with masked regions containing extremely bright discrete sources, we fill each of them with the mean
brightness temperature value calculated from the surrounding pixels after the raw separation is finished.

\subsubsection{Identification of Bright Galaxy Clusters}
In order to locate galaxy clusters on the raw separated images $\bar{T}{_{\rm B,~65{\rm MHz}}^{\rm CL}}^{ \prime}({\bf r})$, we introduce a detection algorithm based on the wavelet transform (WT; Slezak et al. 1990),
which enable us to discriminate structures as a function of scale,
especially those small-scale ones embedded within the
larger ones. To be specific, we adopt the \`{a} Trous wavelet transform algorithm (Shensa 1992; Vikhlinin et al. 1997; Gu et al. 2009), which is a nonorthogonal version of discrete wavelet transform. 
 We employ a Gaussian-shaped
convolution mask $h(\sigma,l) = e^{-l^2/(2\sigma^2)}/(2\pi\sigma^2)$,
and define the corresponding wavelet kernel as $W(l) =
h(\sigma,l)-h(2\sigma,l)$, where $\sigma$ is the scaling
parameter and $l=2^n$ pixels ($n=1, ~2,~\ldots$) is the scale. For each raw separated image $\bar{T}{_{\rm B,~65{\rm MHz}}^{{\rm CL}}}^{ \prime}({\bf r})$, we apply the
convolution mask to the map  to obtain a smooth plane for the scale
of $l=2^{1}$ pixels. Then, we apply the convolution mask to the obtained
smooth plane to obtain a smooth plane for the next scale
(i.e., $l=2^{2}$ pixels). After repeating this process, a set of six smooth
planes $c_{i}(x,y)$ ($i=1,~2,~\ldots,~6$) are obtained for scales of
$l=2^{n}$ ($n=1,~2,~\ldots,~6$) pixels, respectively. By calculating the
difference between any two smooth planes that have adjacent
scales, we obtain five subimages
$w_{i}(x,y)=c_{i}(x,y)-c_{i+1}(x,y)$, $i=1,~2,~\ldots,~5$. 
 The detection of clusters is performed on the subimages with the 3th and 4th scales, which are larger than the sizes of discrete sources convolved by the 21CMA PSF.

When the strength of the ICM magnetic field is set to be $B=2 ~\mu{\rm G}$, 47 cluster-like structures are detected on the raw  separated image $\bar{T}{_{\rm B,~65{\rm MHz}}^{{\rm CL}}}^{ \prime}({\bf r})$, of which 37 are confirmed to be simulated clusters and 10 be simulated discrete sources,
corresponding to a trust ratio of $N{\rm^{CL}_{detect}}/N{\rm^{Total}_{detect}}=37/47 \simeq 79\%$. We also have examined the central
brightness temperature $\bar{T}^{\rm CL}_{0,~65{\rm MHz}}$ of the 37 detected clusters, and find that all of
them have  $\bar{T}^{\rm CL}_{0,~65{\rm MHz}}>10$ K,  corresponding to a completeness ratio of $N{\rm^{CL}_{detect}}/N{\rm^{CL}_{simulate}}|_{\bar{T}^{\rm CL}_{0,~65{\rm MHz}}>10~{\rm K}}=37/48 \simeq 77\%$ for all simulated clusters that have $\bar{T}^{\rm CL}_{0,~65{\rm MHz}}>10~{\rm K}$ (Fig.   7{\it c}). When $B=0.2 ~\mu{\rm G}$ is assumed for the ICM, 19 cluster-like structures (15 simulated clusters and 4 simulated discrete sources) are detected, so that we have $N{\rm^{CL}_{detect}}/N{\rm^{Total}_{detect}} =15/19\simeq 79\%$ and $N{\rm^{CL}_{detect}}/N{\rm^{CL}_{simulate}}|_{\bar{T}^{\rm CL}_{0,~65{\rm MHz}}>10~{\rm K}}=15/18 \simeq 83\%$, respectively (Fig.   7{\it d}). Discrete sources that are misidentified as galaxy clusters at this step can be recognized by examining their  brightness temperature profiles and spectra (see details in \S4.1).  

\section{DISCUSSION}

\subsection{Brightness Temperature and Spectral Information of Detected Bright Clusters}
In order to examine how accurate the FastICA + WT separation algorithm is, we randomly
select three identified clusters with different redshifts and brightness temperatures (Table 2) to extract and study their radial brightness temperature profiles and spectra, and compare them with the input model, under the assumptions of $B=2~\mu$G and $0.2~\mu$G for the ICM, respectively.  
To correctly obtain a cluster's image in the 65 MHz channel ($\bar{T}{_{\rm B,~65{\rm MHz}}^{{\rm CL}}}^{ \prime}({\bf r})$), here we define a local circular region, which is large enough to cover the whole cluster (Fig.    7$c$ \& 7$d$), and feed the fake 21CMA survey maps extracted in this local region in five adjacent channels ($\nu=$61, 63, 65, 67 and 69 MHz, $\Delta \nu=1$ MHz for each channel) into the FastICA program. The separated images (Fig.   8) are obtained with the same  FastICA + WT algorithm and are rescaled in the same way as described in $\S{\it 3.2.1}$.   By repeating this process, we obtain separated images of each cluster in the  80, 95, 110, 125, 140, 155, 170, and 185 MHz ($\Delta \nu=1$ MHz for each channel) channels, respectively.
Within a radius at which the brightness temperature drops to 1/10 of its
central value, we extract the brightness temperature profiles and spectra of each cluster from the separated  images (Fig.    9).  At a given radius $R$ as measured from the cluster's center ${\bf r_0}$, the systematic error is calculated as the mean deviation 
$
\Delta{\bar{T}{_{\rm B,~\nu}^{{\rm CL}}}^{ \prime}}(R) =\sqrt{{\sum_{|{\bf r}-{\bf r}_0|=R} [\bar{T}{_{\rm B,~
\nu}^{\rm CL}}^{ \prime}({\bf r})-<{\bar{T}{}_{\rm B,~\nu}^{\rm CL}}^{ \prime}(R)>]^2}/{N_{\rm pixel}|_{{|{\bf r}-{\bf r}_0|=R} }}}
$,
where $<{\bar{T}{}_{\rm B,~\nu}^{\rm CL}}^{ \prime}(R)>$ is the brightness temperature averaged at $R$, and $N_{\rm pixel}|_{|{\bf r}-{\bf r}_0|=R} $ is the area.

In Figure 9, we also show the original input radio brightness temperature profiles and spectra of the three selected clusters, as well as the difference between the extracted and input distributions as residuals.
In each case we find that the extracted and input distributions agree very well with each other within the error bars.    
Consequently, we conclude that the FastICA  + wavelet algorithm will enable us to separate and study the brightest ($\bar{T}^{\rm CL}_{0,~65{\rm MHz}}>10~{\rm K}$) clusters in the 21CMA FOV, whose number is expected to be about 37 if $B=2~\mu$G (or 15 if $B=0.2~\mu$G).

  At 65 MHz, we successfully identify about $80\%$ (37 out of 48 if $B=2~\mu$G, or 15 out of 18 if $B=0.2~\mu$G) of the simulated clusters that possess a central brightness temperature of $\bar{T}^{\rm CL}_{0,~65{\rm MHz}}=10- 5000 ~{\rm K}$, comparing to the average  brightness temperature of $\simeq 2500~{\rm K}$ of the foreground Galactic emission. The detected clusters are distributed in a mass range of $2\times10^{14}~M_{\odot}-1.5\times10^{15}~M_{\odot}$, and a redshift range of $z=0.02-1.6$ if $B=2~\mu$G, or $z=0.02-0.8$ if $B=0.2~\mu$G. All of them are bright enough for a direct model study to determine their spectral properties.  We ascribe the failure of  detection of the remaining 14 clusters (11 clusters if $B=2~\mu$G, or three clusters if $B=0.2~\mu$G) to the following three reasons. First, six out of the 14 undetected clusters exhibit relatively flat spectra, which makes it difficult to disentangle them from the Galaxy component, the spectral index of which ranges from $\simeq 2.7$ to 2.9. Second, five out of the 14 clusters are faint and less extended, either embeded in brighter clusters or overlapped by bright discrete sources. And third, three out of the 14 clusters are less extended and are masked along with adjacent bright discrete sources.
We also find that 10  (if $B=2~\mu$G) or four  discrete sources (if $B=0.2~\mu$G) are misidentified as galaxy clusters. This mistake can be corrected by examining the  brightness temperature profiles and spectra of the sources.  

For the cosmological 21cm signals survey projects such as the 21CMA, although our separation approach can only extract  bright clusters, which are more than two orders of magnitude brighter than the cosmological 21cm signals, from the images,  the analysis on the cluster component can deepen our understanding on cluster characters  (e.g., brightness, shape, spatial distribution, spectrum) in the low-frequency radio sky, the information of which is valuable when detecting the  cosmological 21cm signals in both image and frequency spaces.

\subsection{Existence Rate of Cluster Radio Halos}
In the frame of the most popular model for cluster radio halos, i.e., the turbulent re-acceleration model (e.g., Ferrari et al. 2008), every cluster should host a radio halo after experiencing a major merger. However, since the halo's radio emission decays quickly as the relativistic electrons lose their energy, radio halos are expected in only about $20-30\%$ of massive ($M^{\rm CL} \gtrsim 2\times 10^{15}~M_{\odot}$) clusters and in $2-5\%$ of less massive ($M^{\rm CL} \simeq 10^{15}~M_{\odot}$) clusters (Cassano et al. 2006), in other words,  in the whole sky $\lesssim$ 50 clusters  are found to host a bright radio halo at 1.4 GHz according to the obsevational works (Cassano et al. 2006; Giovannini et al. 2009; Brunetti et al. 2009). Using this halo number density we estimate that $\lesssim 1$ radio halo shall appear at 1.4 GHz in the 21CMA FOV, which is consistent with the 1.4 GHz detection of Abell 2294 in this field. On the other hand, by calculating the evolution of the  power density $P_{1.4~{\rm GHz}}^{\rm CL}(t_E)$ of cluster radio halos (see Appendix), we find that among  1084 simulated clusters only one (if $B=2~\mu$G) or none (if $B=0.2~\mu$G) cluster (Fig.  10) is brighter than the standard VLA detection threshold at 1.4 GHz, given an intergration time of  $\simeq 3$ hrs (Giovannini et al. 2009). This shows that our work is nicely consistent with current observations performed at higher frequencies, and the prediction for the apperence of more than one dozen low-frequency cluster radio halos in the 21CMA FOV is reliable.

This work is accomplished based on the predictions of the standard turbulent re-acceleration model, which assumes that most cluster radio halos are formed in major merger events. However, in alternative models (e.g., the magneto-hydrodynamic (MHD) model; Brunetti et al. 2004; Cassano 2009), some radio halos, although relatively small and weak, can also be produced on cluster scales via minor mergers, so that much more halos are expected in the low-frequency radio sky. The survey of galaxy clusters with upcoming facilities working in the band of a few hundred of MHz will help distinguish  these models.

\section{SUMMARY}
By carrying out Monte-Carlo simulations we model the foreground of the cosmological reionization signals, including emissions from our Galaxy, galaxy clusters, and extragalactic discrete sources  (i.e., star-forming galaxies, radio-quiet AGNs, and radio-loud AGNs).  As an improvement of previous works, we consider in detail not only random variations of morphological and spectroscopic parameters within the ranges allowed by  multi-band observations, but also evolution of radio halos in galaxy clusters, assuming that relativistic electrons are re-accelerated  in the ICM in merger events and lose energy via both synchrotron emission and inverse Compton scattering with CMB photons. With these we create the fake $50-200$ MHz sky as will be observed with the 21CMA array. We show that about $80\%$  (37 out of 48 if $B=2~\mu$G, or 15 out of 18 if $B=0.2~\mu$G) of the clusters with a central  brightness temperature of  $> 10~{\rm K}$ at 65 MHz can be successfully disentangled from the fake sky maps, if an ICA+wavelet detection  algorithm is applied.

\begin{acknowledgements} 
We thank the referee for the valuable comments on our manuscript. This work was supported by the National Science Foundation of China (Grant No. 10673008, 10878001 and 10973010), the Ministry of Science and Technology of China (Grant No. 2009CB824900, 2009CB824904), and the Ministry of Education of China (the NCET Program).
\end{acknowledgements}

\clearpage
\section*{Appendix: Evolution of Synchrotron Emission Spectra of Cluster Radio Halos }

By using the approach presented in You (1998; p201), we calculate the evolution of  synchrotron
emission spectrum  for a population of relativistic electrons that lose energy via synchrotron radiation and inverse Compton scattering with CMB photons. 
Assuming that a group of relativistic electrons  are injected into the magnetic field at the initial time (i.e., $t=0$) once only, the number of which possesses an energy distribution of  
\begin{equation}
 N(\gamma,t)|_{t=0}=N_0\gamma_0^{-\Gamma},
\end{equation}
where $N_0$ is the density normalization, $\gamma$ is the Lorentz factor so that the electron energy is $\gamma m_e c^2$, $\gamma_0=\gamma|_{t=0}$ is the  initial electron energy, and $\Gamma$ (defined to be $3$ in this work)
is the initial energy spectral index. Due to the conservation of electron number, 
 electrons distributed within the energy interval $\gamma_0 \sim \gamma_0+ d\gamma_0$ at $t=0$ shall fall in the energy interval $\gamma \sim \gamma+ d\gamma$ at a later time time $t$,  we have
\begin{equation}
 N(\gamma,t)d\gamma=N_0\gamma_0^{-\Gamma}d\gamma_0,
\end{equation}
 where $N(\gamma,t)$ is the  electron number per unit volume at time $t$ within the energy interval $\gamma \sim \gamma+d\gamma$. 
We also know that  electrons lose energy via synchrotron radiation and inverse Compton scattering with CMB photons at a rate of
\begin{equation} 
\dot{\gamma}=-b\gamma^2,
\end{equation}
 where $b=3\times10^8U_{\rm mag}$, and $U_{\rm mag}=[ B^2+(B^{\rm CMB})^2  ]/{8\pi}$ ($B$
is the local magnetic field strength of the ICM, and $B^{\rm CMB}$ is the equivalent
magnetic field strength due to  the inverse Compton scattering between  relativistic electrons and  CMB photons) is the energy density of the synthesized magnetic field (see also Parma et al. 1999; Ferrari et al. 2008),
by solving which we obtain 
\begin{equation}
 \gamma= \frac {\gamma_0 } {1+b\gamma_0t},
\end{equation}
or
\begin{equation}
 \gamma_0=\frac {\gamma}{1-b\gamma t},
\end{equation}
and 
\begin{equation}
d\gamma_0=\frac {1} {(1-b\gamma t)^2}d\gamma. 
\end{equation}
Then by substituting equation (24) and equation (25) into equation (21), we have
\begin{equation}
  N(\gamma,t)d\gamma \\
=\left\{
\begin{array}{ll}
 N_0\gamma^{-\Gamma}(1-b\gamma t)^{\Gamma-2}d\gamma, & \gamma_1<\gamma<\gamma_2, \\
 0,  & \gamma<\gamma_1~ {\rm or}~ \gamma>\gamma_2,\\
\end{array}
\right.
\end{equation} 
where $\gamma_1$ and $\gamma_2$ are the lower and upper limits of electron energy at time $t$, respectively.

On the other hand, if the synchrotron radiation of the electron population is isotropic, the synchrotron
 emission coefficient (the energy emitted per unit time per unit volume per unit frequency) can be simply expressed as   
\begin{equation}
  j_{\nu}(t)=4\pi\int N(\gamma,t)P_{\nu}^{\rm e}d\gamma,
\end{equation}
where $P_{\nu}^{\rm e}\propto
\nu_L\frac{\nu}{\nu_c}\int_{\nu/\nu_c}^\infty K_{5/3}(x)dx$ is the
power density of one electron's emission, $\nu_L=\frac {eB} {m_0c}$
is the Larmor frequency, $\nu_c$ is the critical frequency, and
$K_{5/3}$ is the modified Bessel function of order $5/3$.
Therefore, the power density (the energy emitted per unit time per unit frequency) at any frequency $\nu$ and time $t$ can be calculated with both equation (26) and equation (27), 
\begin{equation}
  P_{\nu}^{\rm CL}(t)=j_{\nu}(t)V=j_{\nu}(t)\frac {P_{\rm 1.4~GHz}^{\rm CL}(t=0)}{j_{1.4~{\rm GHz}}(t=0)},
\end{equation}
where $V$ is the volume of the cluster radio halo, and the initial power density at 1.4 GHz, i.e., $P_{1.4{\rm ~GHz}}^{\rm CL}(t=0)$, has been derived from the X-ray temperature of the cluster (\S{\it 2.2.2}).


\clearpage
\begin{table}[]

\footnotesize
\renewcommand{\thefootnote}{}
  \caption{ Estimated numbers of extragalactic discrete sources and their structural subcomponents  in the 21CMA FOV (\S2.3)}
  \label{Tab:publ-works}

  \begin{center}\begin{tabular}{ccc}
 \hline\hline\noalign{\smallskip}
 Source type & Number of sources ($10^6$) & Number of Structures ($10^6$) \\
\noalign{\smallskip}
  \hline\noalign{\smallskip}
Radio-quiet AGN & 7.1 & 7.1 \\
FR I & 4.7 & 14.1  \\
FR II & 0.0005 & 0.0025  \\
Normal galaxies & 40.8 & 40.8  \\
Starburst galaxies & 1.4& 1.4  \\
GPS \& CSS & 0.4& 0.4  \\ 
  \noalign{\smallskip}\hline
  \end{tabular}\end{center}

\end{table}


\begin{table}[]

\tiny
\renewcommand{\thefootnote}{}
  \caption{Confusion level estimation with different methods (\S2.4)$^{\dag}$}
  \label{Tab:publ-works}

  \begin{center}

\begin{tabular}{cccccc}
\hline \hline\noalign{\smallskip}

Method&&65 MHz&&\multicolumn{2}{c}{150 MHz}\\
 \cline{5-6}\noalign{\smallskip}
&&21CMA Instrument \& Our Image (PSF=$3.2^{\prime}$)&&21CMA Instrument (PSF=$1.6^{\prime}$)&Our Image (PSF=$3.2^{\prime}$)\\
\hline
 \noalign{\smallskip}
I&Our flux density function&36.1 mJy/407 K&&3.33 mJy/28.2 K&18.4 mJy/39.2 K\\
&Literature flux density function&33.1 mJy/373 K&&3.20 mJy/27.1 K&18.2 mJy/38.8 K\\
\hline \noalign{\smallskip}
II&Photometric Criterion& 32.0 mJy/361 K&&3.65 mJy/30.9 K&18.6 mJy/39.6 K\\
\hline \noalign{\smallskip}
III&SExtractor&28.1 mJy/317 K&&3.02 mJy/25.6 K&14.1 mJy/30.0 K\\

  \hline
  \end{tabular}

\end{center}
{$^{\dag}$The confusion level given are $3\sigma$ values, corresponding values of brightness temperature are also listed.}
\end{table}


\begin{table}[]

\footnotesize
\renewcommand{\thefootnote}{}
  \caption{Properties of the randomly selected three clusters, which are separated from the fake 21CMA field (\S4.1)$^{\ddag}$ }
  \label{Tab:publ-works}

  \begin{center}

\begin{tabular}{ccccccccc}
\hline \hline\noalign{\smallskip}

Cluster ID &$M^{\rm CL}$&$z$&$r_{200} $&$T_{\rm X}$  &$P_{1.4~ {\rm GHz}}^{\rm CL}(t=0)$&$t_E$&\multicolumn{2}{c}{$\bar{T}^{\rm CL}_{0,~65{\rm MHz}}$}\\
&$(10^{14}~{\rm M_{\odot}})$&& ($h_{71}^{-1}$ Mpc)&(keV)& ($10^{23}$ W Hz$^{-1}$)&  (Gyr)& \multicolumn{2}{c}{(K)}\\
\cline{8-9}
&&&&&&&$B=2$ $\mu$G&$B=0.2$ $\mu$G\\

\hline\noalign{\smallskip}
1&3.02  &0.07   &1.35	&3.7	&4.37   &0.24   &636.0	&129.2\\
2&3.04	&0.25   &1.27   &3.9    &5.09	&0.11	&146.0	&32.5\\
3&8.49	&0.27	&1.78	&7.0	&29.0	&0.11	&634.1	&111.9\\

  \noalign{\smallskip}\hline
  \end{tabular}\end{center}
{$^{\ddag}$Cluster identifications, masses, redshifts,  virial radii, X-ray temperatures, initial power densities  at 1.4 GHz, evolving time, and central brightness temperatures at 65 MHz (if $B=2$ $\mu$G or $0.2$ $\mu$G) are listed, respectively. }
\end{table}

\begin{figure}[]
     \begin{center}
\includegraphics[width=\textwidth]{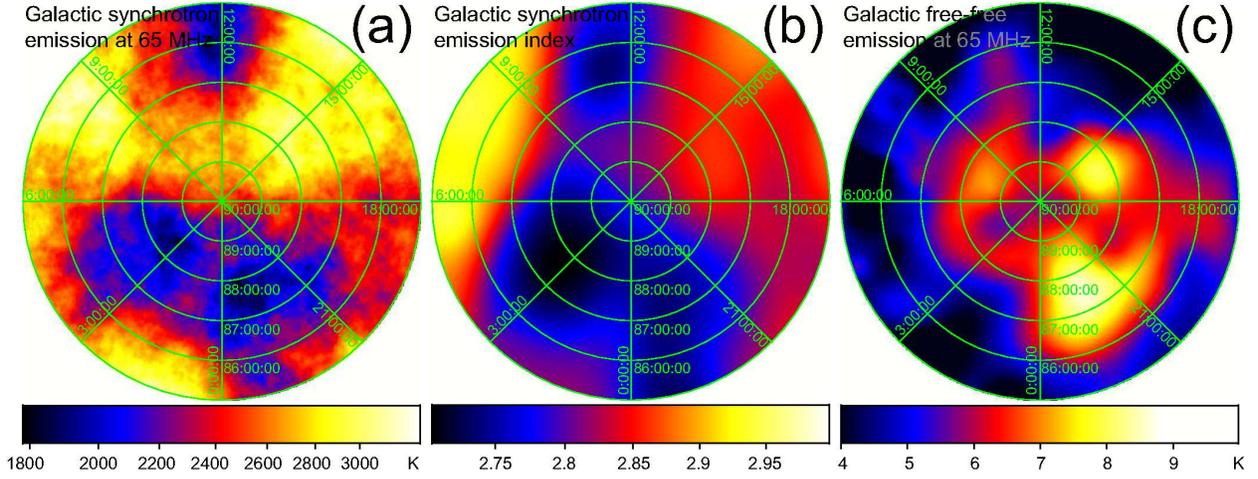}
   \caption{ Simulated two-dimensional images for the distributions of (a)  brightness temperature of the Galactic synchrotron emission at 65 MHz, (b)  Galactic synchrotron  temperature spectral index ($\alpha_{T}^{G,{\rm syn}}$) , and (c)  brightness temperature of the Galactic free-free emission at 65 MHz, all constrained in the 21CMA FOV.
}
   \end{center}
\end{figure}

\begin{figure}[]
     \begin{center}
\includegraphics[width=0.6\textwidth]{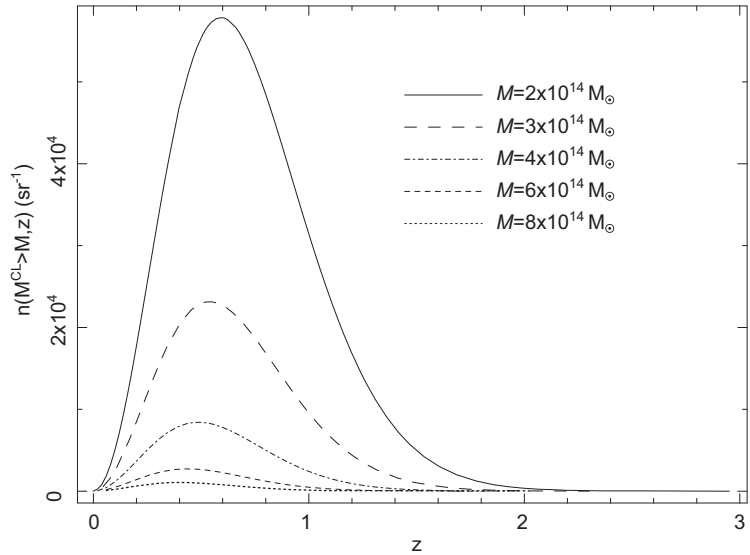}
   \caption{The theoretical number density distribution $n(M^{\rm CL}>M,z)$ (count per unit redshift per unit solid angle) of galaxy clusters as a function of mass and redshift according to the Press-Schechter model. In our simulation, we define $n^{\rm CL}(z)=n(M^{\rm CL}>2\times10^{14}~M_{\odot},z)$.}
   \end{center}
\end{figure}

\begin{figure}[]
     \begin{center}
\includegraphics[width=\textwidth]{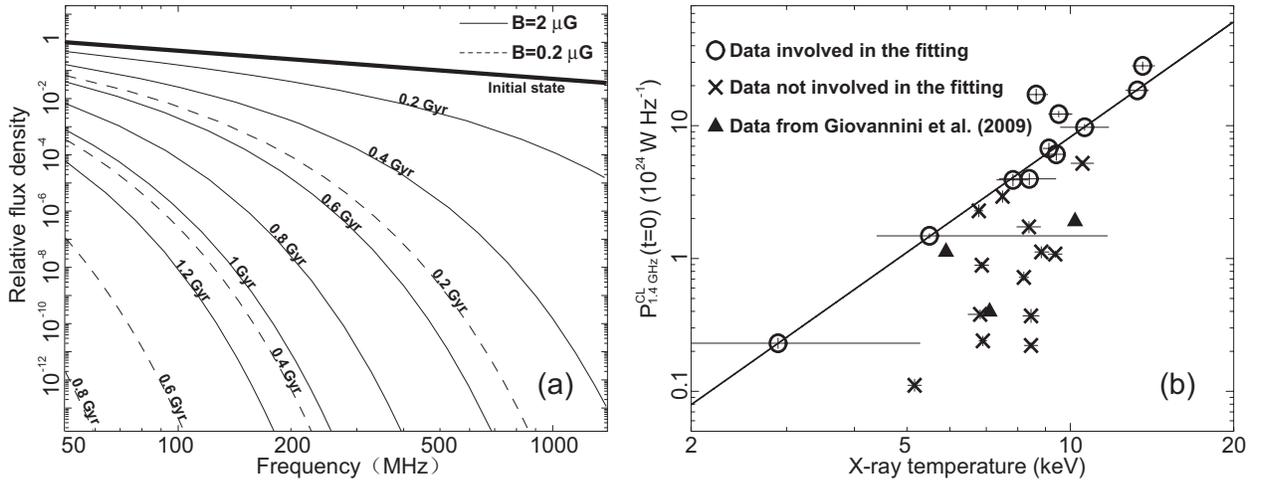}
   \caption{(a) Spectral evolution of  cluster radio halos at $z=0.2$, assuming $B=2$ $\mu$G and $B=0.2$ $\mu$G, respectively. (b) An approximation to the relation between $P_{1.4~{\rm GHz}}^{\rm CL}(t=0)$ and $T_{\rm X}$ (solid line), which is obtained by repeatedly fitting the observed data with maximum likelihood method and excluding the data located outside the $3\sigma$ lower limit (\S{\it 2.2.2}). The marks ``$\circ$''s and ``$\times$''s stand for the data finally involved and not involved in the fitting, respectively. We also superpose the data of three new cluster samples provided by Giovannini et al. (2009) as a comparison  (black triangles).
}
   \end{center}
\end{figure}

\begin{figure}[]
     \begin{center}
\includegraphics[width=\textwidth]{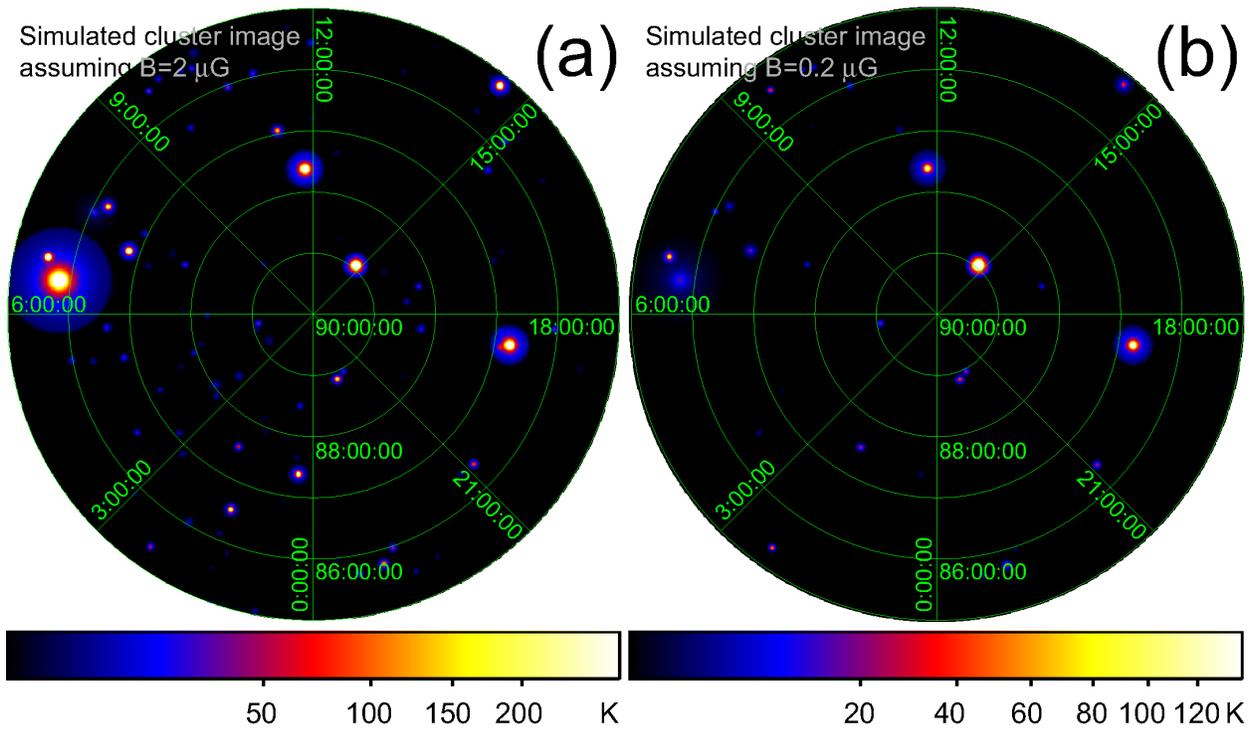}
   \caption{Cluster brightness temperature images simulated at 65 MHz, assuming $B=2$ $\mu$G (a) and $B=0.2$ $\mu$G (b), respectively. Both of the images have been  convolved with a PSF of $3.2^{\prime}$ and are constrained in the 21CMA FOV.}
   \end{center}
\end{figure}

\begin{figure}[]
     \begin{center}
\includegraphics{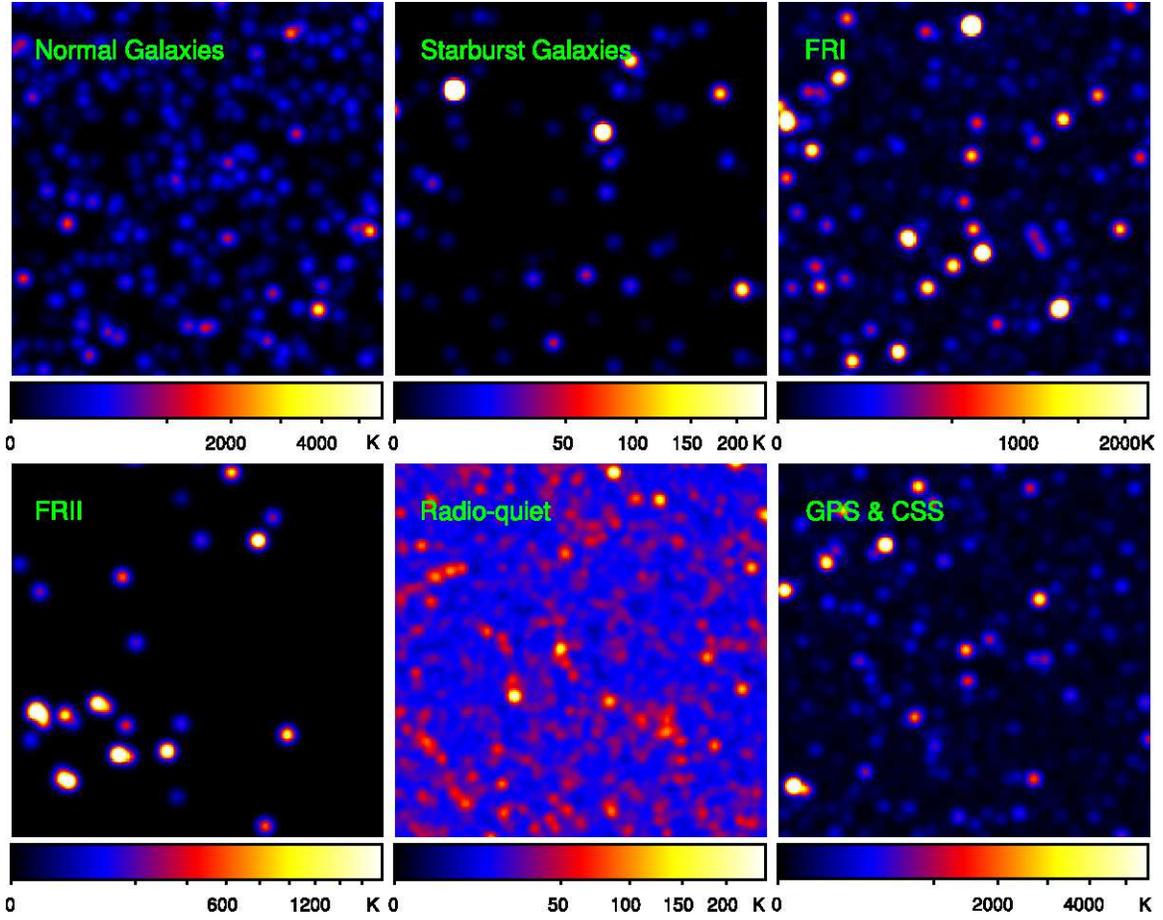}
   \caption{Brightness temperature images of extragalactic discrete sources simulated at 65 MHz, which have been  convolved with a PSF of $3.2^{\prime}$, and are constrained in $2^{\circ}\times2^{\circ}$ fields to illustrate structural details. 
}
   \end{center}
\end{figure}

\begin{figure}[]
     \begin{center}
\includegraphics[width=0.7\textwidth,angle=270]{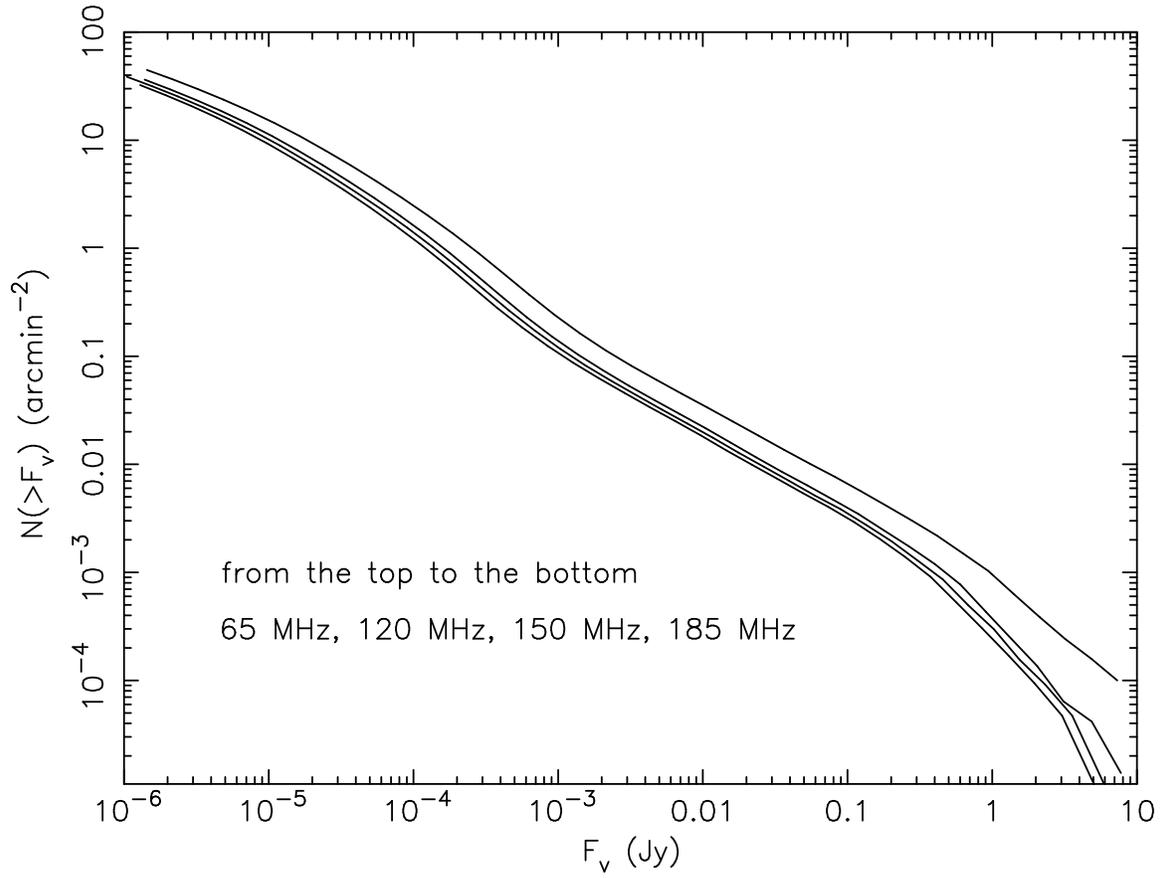}
   \caption{ The total flux density functions of our simulated extragalactic discrete sources.
}
   \end{center}
\end{figure}

\begin{figure}[]
     \begin{center}
\includegraphics[width=\textwidth]{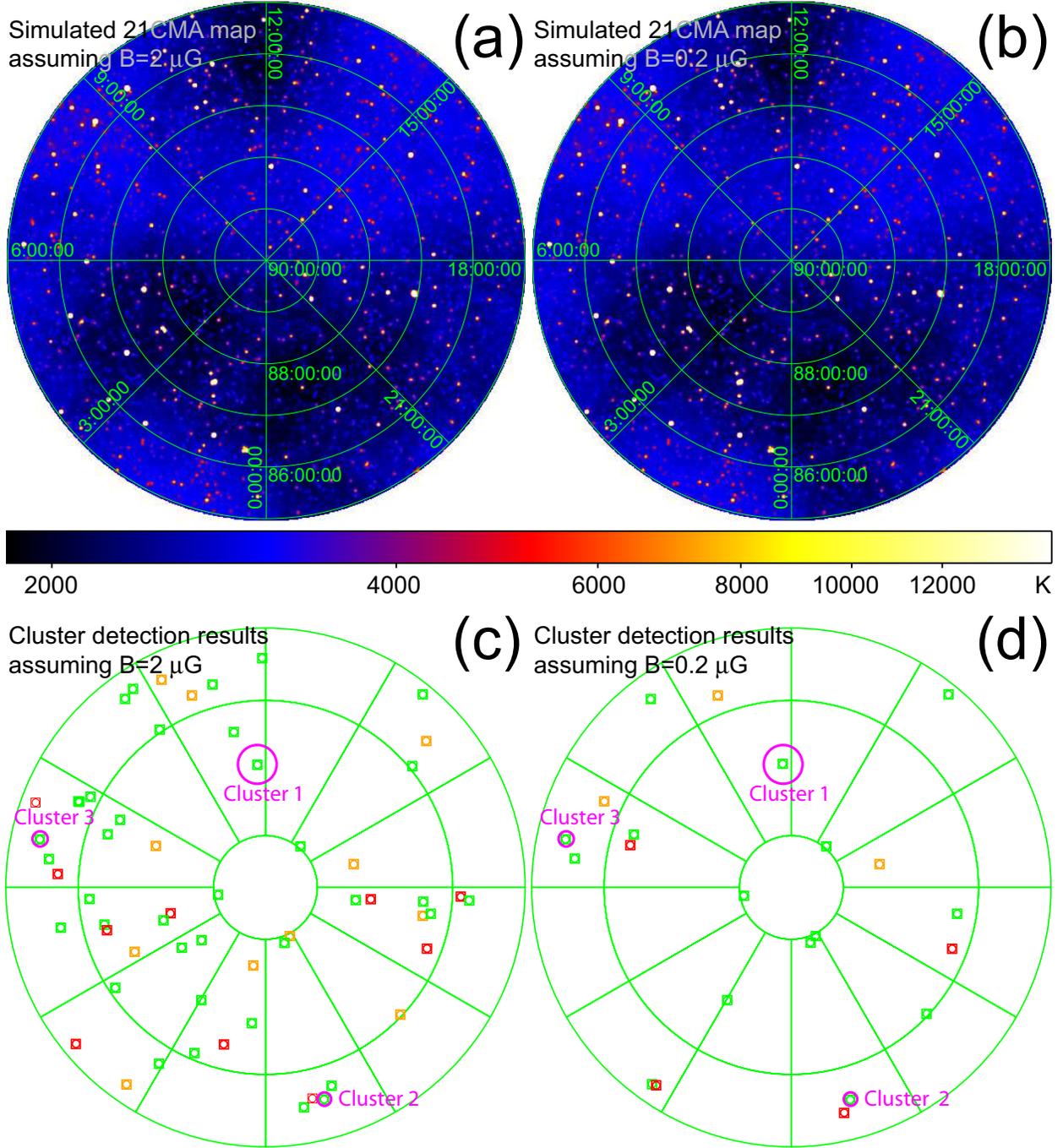}
   \caption{(a)-(b): Simulated 65 MHz 21CMA  maps, assuming $B=2$ $\mu$G and $B=0.2$ $\mu$G, respectively. (c)-(d): Cluster component separated at 65 MHz, where the green squares mark the correctly detected bright ($\bar{T}_{0,65~{\rm MHz}}^{CL} > 10$ K)  clusters, the orange squares mark the missed bright clusters, and the red squares mark the sources misidentified as clusters (\S{\it 3.2.2}). Three magenta circles mark the clusters randomly selected to examine their spatial and spectral properties (\S4.1). 
}
   \end{center}
\end{figure}

\begin{figure}[]
     \begin{center}
\includegraphics[width=\textwidth]{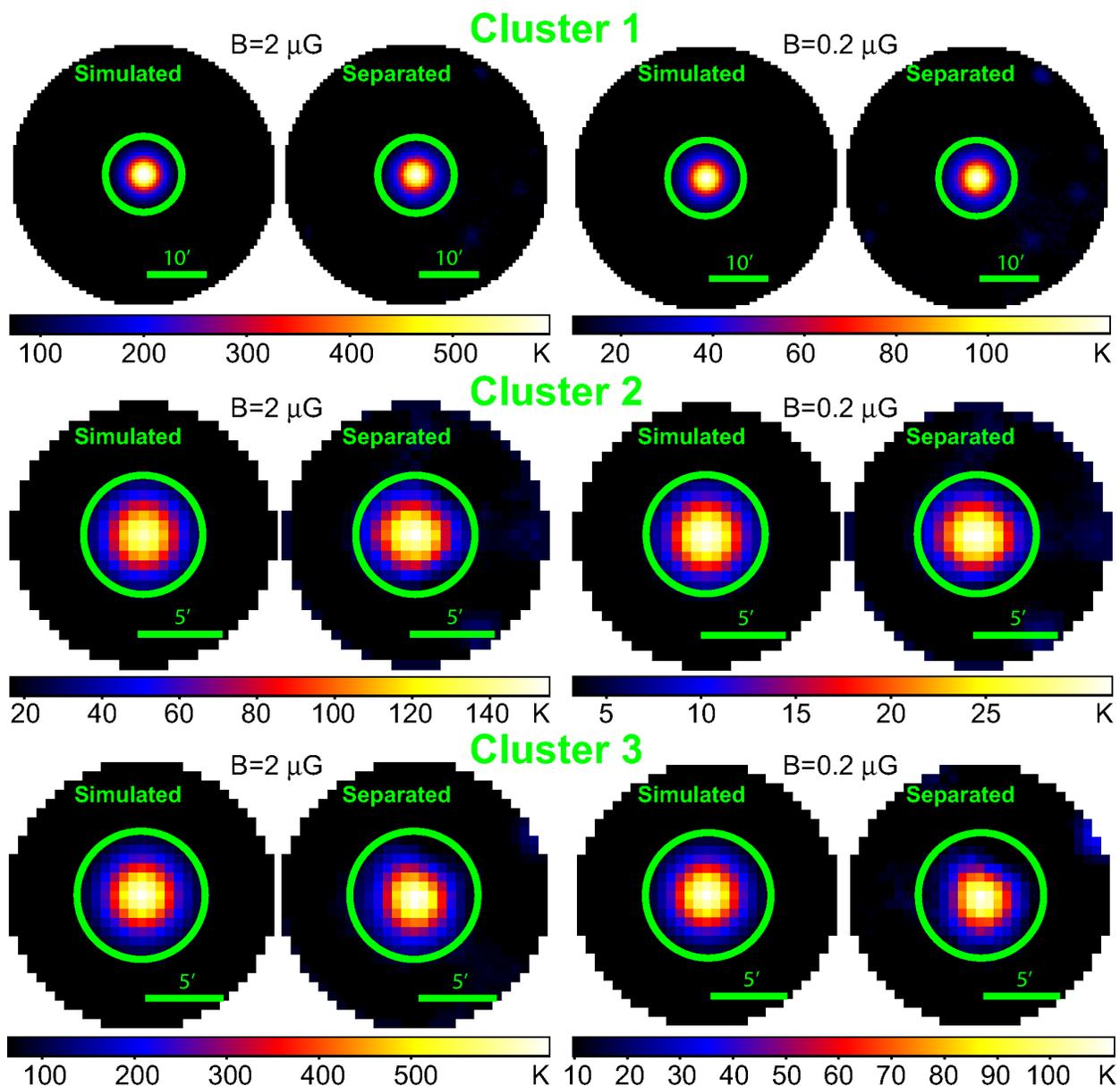}
   \caption{Comparison between the simulated (left panels) and separated (right panels) 65 MHz brightness temperature images of the three randomly selected clusters,  assuming $B=2~\mu$G  and $B=0.2~\mu$G, respectively. The green circles mark the radii at which the brightness temperature drops to 1/10 of the central value.}
   \end{center}
\end{figure}

\begin{figure}[]
     \begin{center}

\includegraphics[width=\textwidth]{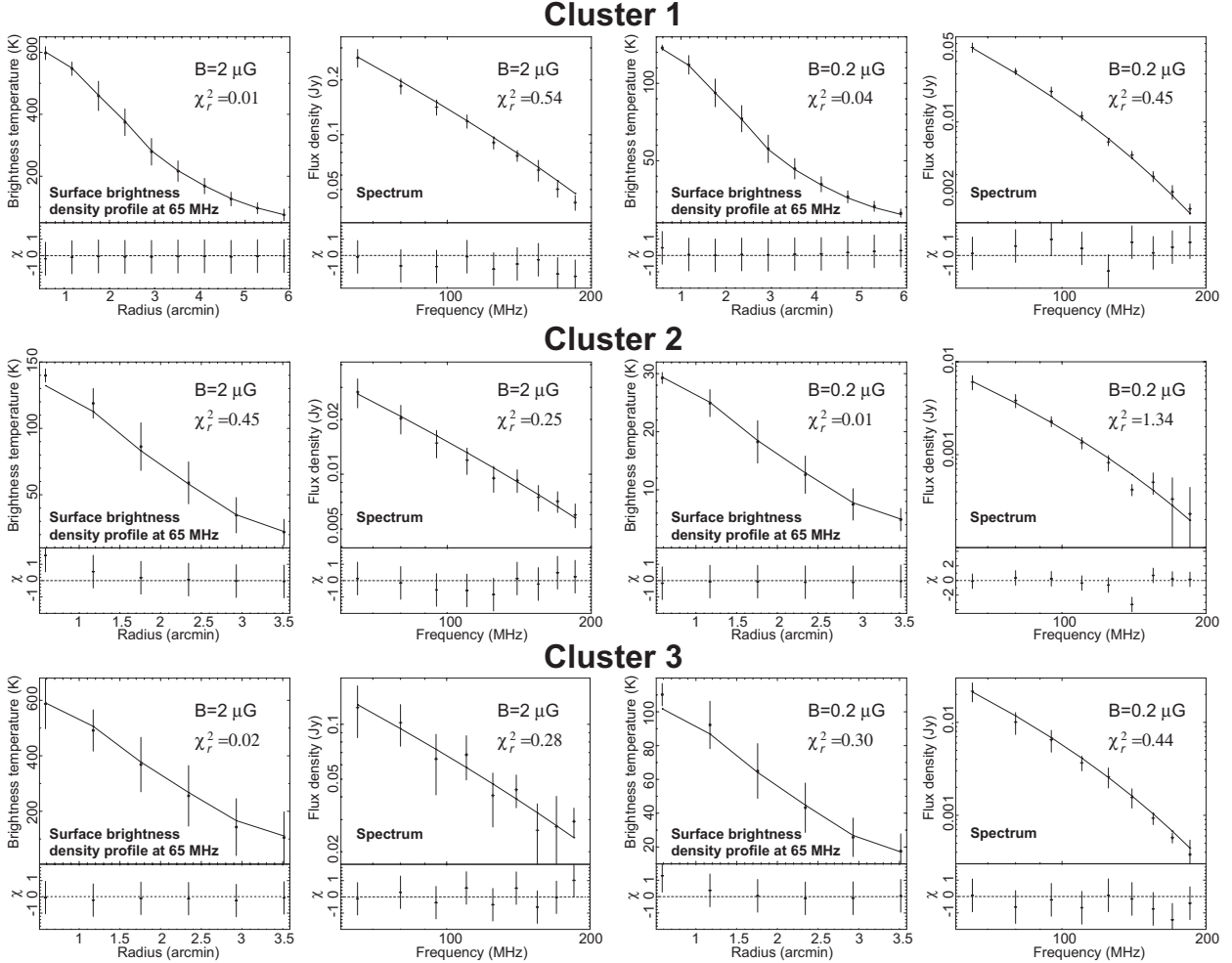}
   \caption{
Brightness temperature profiles in 65 MHz channel and spectra of the three randomly selected clusters extracted within the green circles defined in Fig. 8 (crosses), as compared with the simulated results (solid lines). Residuals obtained in the $\chi^2$ test between the separated distribution and simulated distribution  in each case are also plotted in the lower panels. 
}
   \end{center}
\end{figure} 

\begin{figure}[]
     \begin{center}
\includegraphics[width=\textwidth]{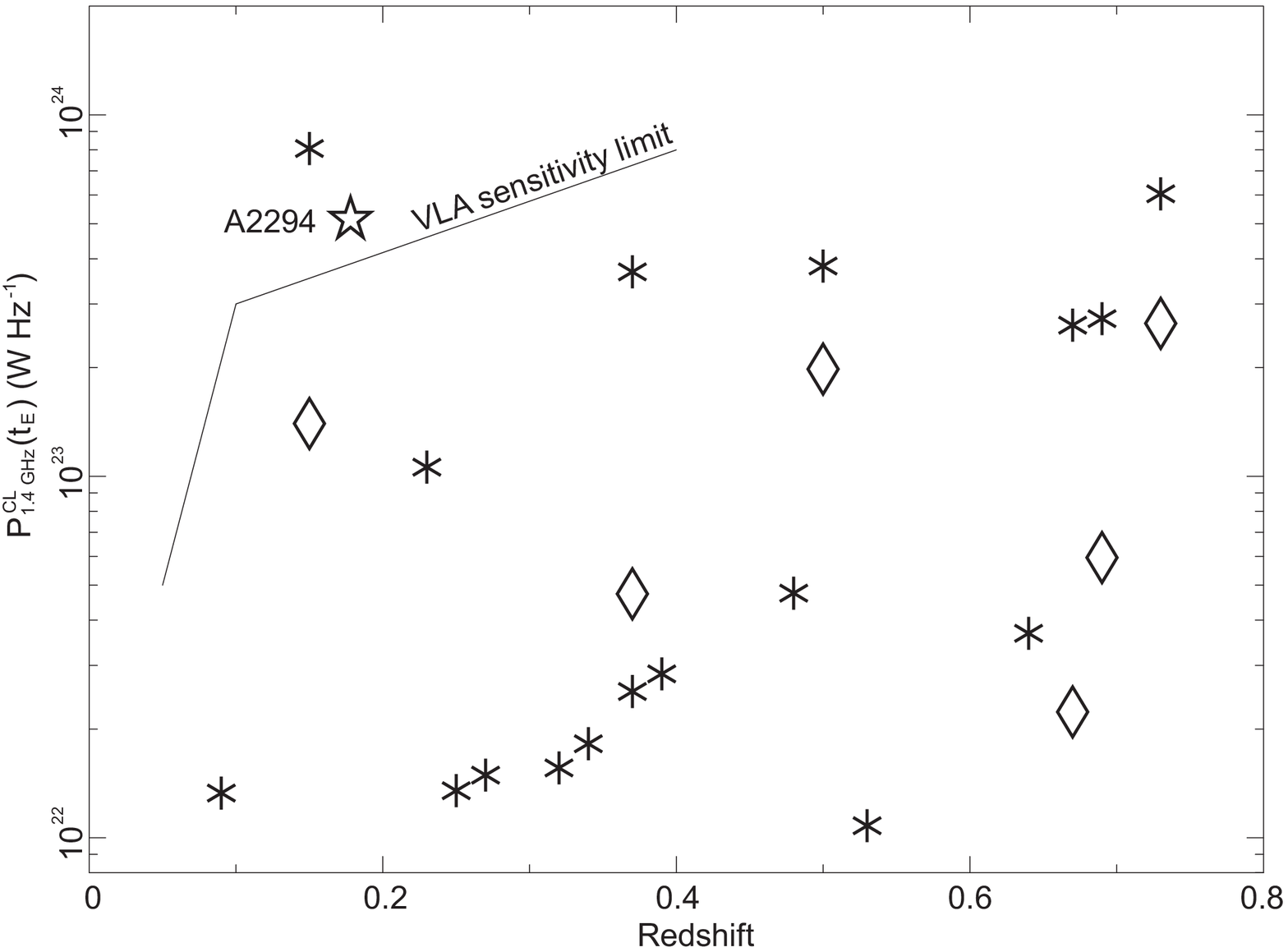}

   \caption{$P_{1.4 {\rm ~GHz}}^{\rm CL}(t_E)-z$ distribution of the simulated clusters, assuming $B=2$ $\mu$G (``$\ast$''s) and $B=0.2$ $\mu$ G (``$\diamondsuit$''s), respectively, as compared with the VLA sensitivity limit,  assuming a standard  observation with an integration time of $\simeq 3$ hrs (Giovannini et al. 2009).  The   star marks  Abell 2294, which is the only one detected cluster that holds a radio halo in the 21CMA FOV nowadays.    }
   \end{center}
\end{figure} 

\clearpage
\begin{references}
\reference{} Arnaud, M., Pointecouteau, E., \& Pratt, G.~W.\ 2005, \aap, 441, 893 
\reference{} Baccigalupi, C., et al.\ 2000, \mnras, 318, 769 
\reference{} Bennett, C.~L., et al.\ 2003, \apjs, 148, 97 

\reference{} Bertin, E., \& Arnouts, S.\ 1996, A\&A, 117, 393 
\reference{} Bernardi, G., et al.\ 2009, \aap, 500, 965 
\reference{} Bottino, M., Banday, A.~J., \& Maino, D.\ 2008, \mnras, 389, 1190 
\reference{} Brunetti, G., Blasi, P., Cassano, R., \& Gabici, S.\ 2004, \mnras, 350, 1174 
\reference{} Brunetti, G., Cassano, R., Dolag, K., \& Setti, G.\ 2009, \aap, 507, 661 
\reference{} Brunetti, G., et al.\ 2008, \nat, 455, 944
\reference{} Brunetti, G., Setti, G., Feretti, L., \& Giovannini, G.\ 2001, \mnras, 320, 365 
\reference{} Cassano, R.\ 2009, Astronomical Society of the Pacific Conference Series, 407, 223 
\reference{} Cassano, R., \& Brunetti, G.\ 2005, \mnras, 357, 1313 
\reference{} Cassano, R., Brunetti, G., R{\"o}ttgering, H.~J.~A., \& Br{\"u}ggen, M.\ 2010, \aap, 509, A68 
\reference{} Cassano, R., Brunetti, G., \& Setti, G.\ 2006, \mnras, 369, 1577 
\reference{} Cavaliere, A., \& Fusco-Femiano, R.\ 1976, \aap, 49, 137 
\reference{} Comon, P. 1994, Signal Processing, 36, 287

\reference{} De Bernardis, P., Bucher, M., Burigana, C., \& Piccirillo, L.\ 2009, Experimental Astronomy, 23, 5 (See http://www.b-pol.org for a full copy of the B-Pol proposal)

\reference{} de Vries, W.~H., Barthel, P.~D., \& O'Dea, C.~P.\ 1997, \aap, 321, 105 

\reference{} Dole, H., Lagache, G., \& Puget, J.-L.\ 2003, ApJ, 585, 617 
\reference{} Drexler, J.\ 2007, arXiv:physics/0702132 
\reference{} En{\ss}lin, T.~A., \& Gopal-Krishna 2001, \aap, 366, 26 
\reference{} En{\ss}lin, T.~A., \& R\"{o}ttgering, H.\ 2002, \aap, 396, 83 

\reference{} Ettori, S.\ 2005, \mnras, 362, 110 
\reference{} Evrard, A.~E., Metzler, C.~A., \& Navarro, J.~F.\ 1996, \apj, 469, 494 

\reference{} Fanti, C., Pozzi, F., Dallacasa, D., Fanti, R., Gregorini, L., Stanghellini, C., \& Vigotti, M.\ 2001, \aap, 369, 380 

\reference{} Fedeli, C., Meneghetti, M., Bartelmann, M., Dolag, K., \& Moscardini, L.\ 2006, \aap, 447, 419 
\reference{} Feretti, L.\ 2005, X-Ray and Radio Connections (eds.~L.O.~Sjouwerman and K.K Dyer) Published electronically by NRAO, http://www.aoc.nrao.edu/events/xraydio Held 3-6 February 2004 in Santa Fe, New Mexico, USA, (E8.02) 10 pages  
\reference{} Ferrari, C., Govoni, F., Schindler, S., Bykov, A.~M., \& Rephaeli, Y.\ 2008, Space Science Reviews, 134, 93
\reference{} Finkbeiner, D.~P.\ 2003, \apjs, 146, 407 
\reference{} Gaensler, B.~M., Dickey, J.~M., McClure-Griffiths, N.~M., Green, A.~J., Wieringa, M.~H., \& Haynes, R.~F.\ 2001, \apj, 549, 959 
\reference{} Giardino, G., Banday, A.~J., G{\'o}rski, K.~M., Bennett, K., Jonas, J.~L., \& Tauber, J.\ 2002, \aap, 387, 82 
\reference{} Giovannini, G., Bonafede, A., Feretti, L., Govoni, F., Murgia, M., Ferrari, F., \& Monti, G.\ 2009, \aap, 507, 1257 
\reference{} Gleser, L., Nusser, A., \& Benson, A.~J.\ 2008, \mnras, 391, 383 
\reference{} Govoni, F., En{\ss}lin, T.~A., Feretti, L., \& Giovannini, G.\ 2001, \aap, 369, 441 
\reference{} Gray, A.~D., Landecker, T.~L., Dewdney, P.~E., Taylor, A.~R., Willis, A.~G., \& Normandeau, M.\ 1999, \apj, 514, 221 
\reference{} Gu, L., et al.\ 2009, \apj, 700, 1161 
\reference{} Haslam, C.~G.~T., Salter, C.~J., Stoffel, H., \& Wilson, W.~E.\ 1982, \aaps, 47, 1 
\reference{} Haverkorn, M., Katgert, P., \& de Bruyn, A.~G.\ 2000, \aap, 356, L13 
\reference{} Holder, G., Haiman, Z., \& Mohr, J.~J.\ 2001, \apjl, 560, L111 
\reference{} Hyv\"{a}rinen A., 1999, IEEE Signal Processing Lett. 6, 145
\reference{} Jeli{\'c}, V., et al.\ 2008, \mnras, 389, 1319 
\reference{} Jones, A.~W., Hobson, M.~P., Mukherjee, P., \& Lasenby, A.~N.\ 1998a, arXiv:astro-ph/9810235 
\reference{} Jones, L.~R., Scharf, C., Ebeling, H., Perlman, E., Wegner, G., Malkan, M., \& Horner, D.\ 1998b, \apj, 495, 100 
\reference{} Kim, K.-T., Kronberg, P.~P., Dewdney, P.~E., \& Landecker, T.~L.\ 1990, \apj, 355, 29 
\reference{} Lazarian, A.\ 2006, \apjl, 645, L25
\reference{} Leach, S.~M., et al.\ 2008, \aap, 491, 597 
\reference{} Lumb, D.~H., et al.\ 2004, \aap, 420, 853 
\reference{} Maino, D., Donzelli, S., Banday, A.~J., Stivoli, F., \& Baccigalupi, C.\ 2006, arXiv:astro-ph/0609228 
\reference{} Majumdar, S., \& Mohr, J.~J.\ 2004, \apj, 613, 41 
\reference{} Mao, X.-C., \& Wu, X.-P.\ 2008, \apjl, 673, L107 
\reference{} Mathiesen, B.~F., \& Evrard, A.~E.\ 2001, \apj, 546, 100 
\reference{} Morales, M.~F., \& Hewitt, J.\ 2004, \apj, 615, 7 

\reference{} O'Dea, C.~P.\ 1998, \pasp, 110, 493 

\reference{} Parma, P., Murgia, M., Morganti, R., Capetti, A., de Ruiter, H.~R., \& Fanti, R.\ 1999, \aap, 344, 7 
\reference{} Peacock, J.~A.\ 1999, Cosmological Physics, by John A.~Peacock, pp.~704.~ISBN 052141072X.~Cambridge, UK: Cambridge University Press, January 1999
\reference{} Peterson, J.~B., Pen, U., \& Wu, X.\ 2005, Astronomical Society of the Pacific Conference Series, 345, 441 
\reference{} Platania, P., Burigana, C., Maino, D., Caserini, E., Bersanelli, M., Cappellini, B., \& Mennella, A.\ 2003, \aap, 410, 847 
\reference{} Popesso, P., B{\"o}hringer, H., Brinkmann, J., Voges, W., \& York, D.~G.\ 2004, \aap, 423, 449 
\reference{} Press, W.~H., \& Schechter, P.\ 1974, \apj, 187, 425 
\reference{} Rephaeli, Y., \& Gruber, D.\ 2003, arXiv:astro-ph/0305354
\reference{} Rephaeli, Y., Gruber, D., \& Arieli, Y.\ 2006, \apj, 649, 673 
\reference{} Reynolds, R.~J., \& Haffner, L.~M.\ 2000, arXiv:astro-ph/0010618 
\reference{} Rosati, P., Borgani, S., \& Norman, C.\ 2002, \araa, 40, 539
\reference{} R{\"o}ttgering, H., de Bruyn, A.~G., Fender, R.~P., Kuijpers, J., van Haarlem, M.~P., Johnston-Hollitt, M., \& Miley, G.~K.\ 2003, Texas in Tuscany.~XXI Symposium on Relativistic Astrophysics, 69 
\reference{} R{\"o}ttgering, H.~J.~A., et al. 2006, arXiv:astro-ph/0610596 

\reference{} R{\"o}ttgering, H.~J.~A., et al. 2008, http://www.strw.leidenuniv.nl/lofar/images/Documents/ Outdated/lofar-surveys-specs.pdf


\reference{} Santos, M.~G., Cooray, A., \& Knox, L.\ 2005, \apj, 625, 575 
\reference{} Sarazin, C.~L., \& Kempner, J.~C.\ 2000, \apj, 533, 73 
\reference{} Schekochihin, A.~A., Cowley, S.~C., Kulsrud, R.~M., Hammett, G.~W., \& Sharma, P.\ 2005, \apj, 629, 139 
\reference{} Shaver, P.~A., Windhorst, R.~A., Madau, P., \& de Bruyn, A.~G.\ 1999, \aap, 345, 380 
\reference{} Shensa, M. J. 1992, IEEE Trans. Signal Processing, 40, 2464
\reference{} Slezak, E., Bijaoui, A., \& Mars, G.\ 1990, \aap, 227, 301
\reference{} Smoot, G.~F.\ 1998, arXiv:astro-ph/9801121 

\reference{} Snellen, I.~A.~G., Schilizzi, R.~T., de Bruyn, A.~G., Miley, G.~K., Rengelink, R.~B., R{\"o}ttgering, H.~J., \& Bremer, M.~N.\ 1998, \aaps, 131, 435 

\reference{} Snellen, I.~A.~G., Schilizzi, R.~T., Miley, G.~K., de Bruyn, A.~G., Bremer, M.~N., R{\"o}ttgering, H.~J.~A.\ 2000, \mnras, 319, 445 

\reference{} Spergel, D.~N., et al.\ 2003, \apjs, 148, 175 
\reference{} Stawarz, {\L}., Ostorero, L., Begelman, M.~C., Moderski, R., Kataoka, J., \& Wagner, S.\ 2008, \apj, 680, 911 


\reference{} Sun, X.~H., Reich, W., Waelkens, A., \& En{\ss}lin, T.~A.\ 2008, \aap, 477, 573

\reference{} Sun, X.~H., \& Reich, W.\ 2009, \aap, 507, 1087 
\reference{} Thompson, A.~R., Moran, J.~M., \& Swenson, G.~W., Jr.\ 2001, Interferometry and synthesis in radio astronomy by A.~Richard Thompson, James M.~Moran, and George W.~Swenson, Jr.~2nd ed.~ New York : Wiley, c2001.xxiii, 692 p.~: ill.~; 25 cm.~''A Wiley-Interscience publication.''  Includes bibliographical references and indexes.~ISBN :  0471254924  

\reference{} Tozzi, P., Madau, P., Meiksin, A., \& Rees, M.~J.\ 2000, \apj, 528, 597 
\reference{} Tucci, M., Carretti, E., Cecchini, S., Nicastro, L., Fabbri, R., Gaensler, B.~M., Dickey, J.~M., \& McClure-Griffiths, N.~M.\ 2002, \apj, 579, 607 

\reference{} V{\"a}is{\"a}nen, P., Tollestrup, E.~V., \& Fazio, G.~G.\ 2001, MNRAS, 325, 1241 

\reference{} Venturi, T., Giacintucci, S., Brunetti, G., Cassano, R., Bardelli, S., Dallacasa, D., \& Setti, G.\ 2007, \aap, 463, 937
\reference{} Venturi, T., Giacintucci, S., Dallacasa, D., Cassano, R., Brunetti, G., Bardelli, S., \& Setti, G.\ 2008, \aap, 484, 327 
\reference{} Vikhlinin, A., Forman, W., \& Jones, C.\ 1997, \apjl, 474, L7 
\reference{} Voit, G.~M.\ 2005, Reviews of Modern Physics, 77, 207 
\reference{} Wang, X., Tegmark, M., Santos, M.~G., \& Knox, L.\ 2006, \apj, 650, 529 
\reference{} Wieringa, M.~H., de Bruyn, A.~G., Jansen, D., Brouw, W.~N., \& Katgert, P.\ 1993, \aap, 268, 215 
\reference{} Wilman, R.~J., et al.\ 2008, \mnras, 388, 1335 
\reference{} You, J.~H., Radiation Mechanism in Astrophysics, Edition 2, Beijing: Science Press, 1998, p. 201
\reference{} Zentner, A.~R., Berlind, A.~A., Bullock, J.~S., Kravtsov, A.~V., \& Wechsler, R.~H.\ 2005, \apj, 624, 505 
\reference{} Zhang, Y.-Y., \& Wu, X.-P.\ 2003, \apj, 583, 529
\reference{} Zito, T., Wilbert, N., Wiskott, L., Berkes, P.\ 2009, Frontiers in Neuroinformatics, 2, 8


\end{references}
\end{document}